# Novel Dual-Channel Long Short-Term Memory Compressed Capsule Networks for Emotion Recognition

Ismail Shahin[1,a,*], Noor Hindawi[1,b], Ali Bou Nassif[2,c], Adi Alhudhaif[3,d], Kemal Polat[4,e]

[1]Department of Electrical Engineering, University of Sharjah, Sharjah, UAE, 27272

[2]Department of Computer Engineering, University of Sharjah, Sharjah, UAE, 27272

[3]Department of Computer Science, College of Computer Engineering and Sciences in Al-kharj, Prince Sattam bin Abdulaziz University, P.O. Box 151, Al-Kharj 11942, Saudi Arabia

[4]Bolu Abant Izzet Baysal University, Faculty of Engineering, Department of Electrical and Electronics Engineering, Bolu, Turkey

[a]ismail@sharjah.ac.ae, [b]U18105940@sharjah.ac.ae, [c]anassif@sharjah.ac.ae, [d]a.alhudhaif@psau.edu.sa, [e]kpolat@ibu.edu.tr

*Corresponding Author: Ismail Shahin

## Abstract

Recent analysis on speech emotion recognition (SER) has made considerable advances with the use of MFCC's spectrogram features and the implementation of neural network approaches such as convolutional neural networks (CNNs). The fundamental issue of CNNs is that the spatial information is not recorded in spectrograms. Capsule networks (CapsNet) have gained gratitude as alternatives to CNNs with their larger capacities for hierarchical representation. However, the concealed issue of CapsNet is the compression method that is employed in CNNs cannot be directly utilized in CapsNet. To address these issues, this research introduces a text-independent and speaker-independent SER novel architecture, where a dual-channel long short-term memory compressed-CapsNet (DC-LSTM COMP-CapsNet) algorithm is proposed based on the structural features of CapsNet. Our proposed novel classifier can ensure the energy efficiency of the model and adequate compression method in speech emotion recognition, which is not delivered through the original structure of a CapsNet. Moreover, the grid search (GS) approach is used to attain optimal solutions. Results witnessed an improved performance and reduction in the training and testing running time. The speech datasets used to evaluate our algorithm are: Arabic Emirati-accented corpus, English "speech under simulated and actual stress (SUSAS)" corpus, English Ryerson audio-visual database of emotional speech and song (RAVDESS) corpus, and crowd-sourced emotional multimodal actors dataset (CREMA-D). This work reveals that the optimum feature extraction method compared to other known methods is MFCCs delta-delta. Using the four datasets and the MFCCs delta-delta, DC-LSTM COMP-CapsNet surpasses all the state-of-the-art systems, classical classifiers, CNN, and the original CapsNet. Using the Arabic Emirati-accented corpus, our results demonstrate that the proposed work yields average emotion recognition accuracy of 89.3% compared to 84.7%, 82.2%, 69.8%, 69.2%, 53.8%, 42.6%, and 31.9% based on CapsNet, CNN, support vector machine (SVM), multi-layer perceptron (MLP), k-nearest neighbor (KNN), radial basis function (RBF), and naïve Bayes (NB), respectively.

**Keywords:** Capsule networks; convolutional neural network; deep neural network; dual-channel; emotion recognition; LSTM.

## 1. Introduction

Speech emotion recognition (SER) can be expressed as the extraction of the emotional talking condition of the speaker from his/her speech signal. Essential loudness, pitch, speech intensity, and glottal parameters, and frequency are the prosodic features utilized to model the various emotions (Zhou et al., 2009). Emotional robots and human-robot emotional communication have been extensively created and employed in multiple areas (Z. Liu et al., 2016). One of the crucial capabilities of the emotional robot is emotion recognition, which primarily involves body language emotion recognition (Rattanyu & Mizukawa, 2011), facial expression recognition (Sun et al., 2017), and speech emotion recognition (Song et al., 2016). For human communication with robots effortlessly and harmoniously, it is required to identify human emotion with significant accuracy for the robot. As speech signal is simple to understand, it is broadly utilized for emotion recognition in human-robot communication (Z.-T. Liu et al., 2018).

With recurrent neural network (RNN), hidden Markov model (HMM), Gaussian mixture model (GMM), and convolutional neural networks (CNNs), recent experiments in automatic speech emotion recognition (SER) have been integrated. Neural networks in SER systems accept features developed from the deep learning classifiers and anticipate targets at frame-level. The two challenging issues in SER are the extraction of high-level frame-based feature representations and the creation of utterance-level features. In small frames, the voice signals are essentially stationary (Schuller et al., 2011), (El Ayadi et al., 2011). Several acoustic features derived from short frames (e.g., pitch), are thought to be affected by emotions and may give precise local information that is emotionally important (Fernández-Diaz & Gallardo-Antolin, 2020; Uddin & Nilsson, 2020). These frame-based characteristics are often be described as low-level functions (Mirsamadi et al., 2017). Focused solely on the low-level frame-based features, neural networks are used to generate frame-by-frame neural representations, which are described as high-level frame-based representations of features (J. & I.Tashev, 2015). Nevertheless, identification of emotions at the utterance level involves a global representation of features, which includes both specific local knowledge and emotion-related global characteristics.

Additionally, in the field of deep learning, CNNs have been extremely efficient but have weaknesses in their basic design, making them perform less than expected for certain tasks (Kwabena Patrick et al., 2019). CNN is implemented to low-level features such as pitch and energy for learning high-level features, i.e., outputs of the neural network. Layers near the beginning detect simple features (low-level), and deeper layers will detect other complicated features. To make a final decision, CNN utilizes all these features that have been analyzed. This is where the system shortcomings lay — there is no spatial information utilized elsewhere in a CNN, and the pooling function utilized to link layers is ineffective (J. Bae & Kim, 2018a). It has lately been suggested by Sabour et al. that capsule networks (CapsNet) conquer the limitation of CNNs in collecting spatial information (Sabour et al., 2017).

CapsNet is the outcome of almost ten years of research by Sabour et al. (Sabour et al., 2017). CapsNet, unlike CNNs, is fabricated of a network of neurons that inputs and outputs are vectors instead of two scalar values. A capsule contains a community of neurons that carry vectors for action. The vector length determines the probability of existence of the activity described by the capsule, and the orientation assembles the precise instantiation parameters of the activity, (e.g., translation, pitch, energy, rotation information, etc.). A routing algorithm is utilized to pair the related activity vectors to enable the respective capsules in the upper layer, depending on the activity vectors provided by the lower layer of the capsules.

Too much essential information is lost in the pooling phase of CNNs since only the most sufficiently active neurons are selected for promotion to the next layer. This process is the reason why useful spatial information among the layers gets lost. Therefore, Sabour et al. presented a CapsNet whose fundamental architecture is displayed in Fig. 1 (Sabour et al., 2017). In their algorithm, they assumed that the individual's brain could attain flip and translation invariance in a further sophisticated way than pooling. "Capsule units" are parts in the brain known as neurons, whereas the capsule layer is invented of several ''capsule'' units, which sequentially create a capsule network. Furthermore, Sabour and her colleagues suggested utilizing methods known as "dynamic routing" to resolve this problem. Dynamic routing technique is employed only between vectorized features to substitute pooling tasks. Geoffrey Hinton, one of the most influential people in the field of deep learning, proposed this solution, which encodes spatial details into features while still utilizing dynamic routing (by agreement) (Sabour et al., 2017), (G. E. Hinton et al., 2011).

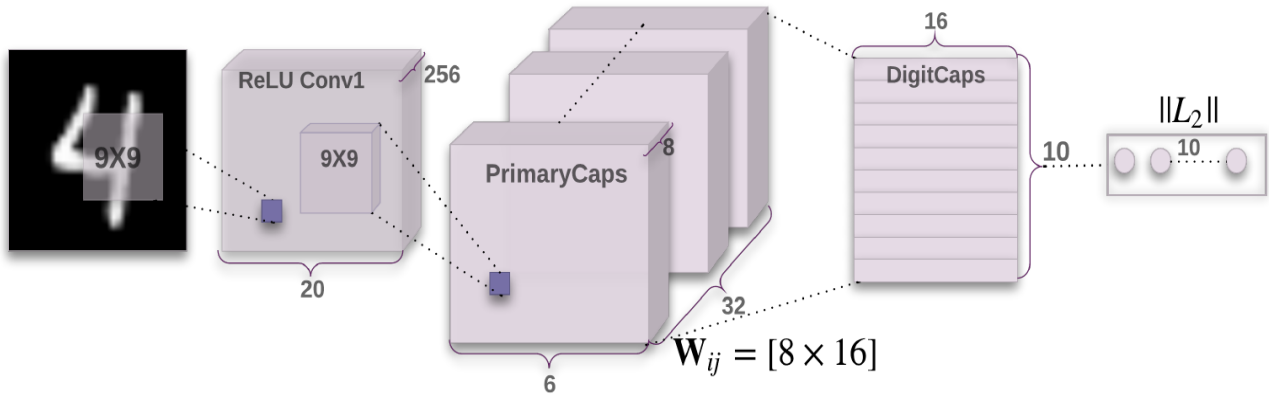

**Fig.1** Basic architecture of CapsNet (Sabour et al., 2017)

In the CapsNet architecture, six main phases are to be completed. In the first phase, the input of the spectrogram or MFCC's representation should be introduced after the preprocessing step. The input is to be inserted into the second phase, where the CNN takes place. Moving to the third phase in the primary capsules, the output of CNNs is the input to these capsules. After that, a dynamic routing mechanism is the fourth phase to allow each capsule at one level to attend to some active capsules at the level below and to ignore others. The fifth phase is called the detection capsule phase, where the main idea of this phase is to detect the correct emotion and organize them each in a row. The final phase is the decoder or Euclidean norm algorithm.

The detailed process of how CapsNet work can be explained as follows:

- A capsule, as mentioned before, is a group of neurons whose activity vector represents the instantiation parameters of a specific type of entity, e.g., an emotion.
- In a capsule network, after having the CNNs, the capsules ordered in the layer are called primary capsules, and each of these capsules in this layer corresponds to a particular entity in the input, e.g., a particular emotion. For example, if the input is a voice, each capsule corresponds to one feature of the voice such as pitch (high, low), frequency, energy, etc.
- Each of these capsules output is a vector, where the length of the vector represents the probability that this feature exists in the voice "P (feature ∈ voice)."
- The next layer is the digital/detection capsules layer, where each capsule in this layer takes information from each capsule in the lower layer and integrates all information.
- In the last step, the correct emotion is recognized.

The main limitation, recently, in CapsNet is the compression method that is employed to CNN, which cannot be directly utilized in the CapsNet. Moreover, CapsNet is a slow algorithm which is due to the inner loop of a dynamic routing algorithm. As the size of the dataset increases, the number of iterations increases. Consequently, CapsNet has higher complexity in implementation compared to CNNs (Jain et al., 2018). Therefore, this study concentrates on utilizing four separate speech datasets to manipulate the capsule nets model for improving SER in emotional talking environments using a newly proposed classifier called dual-channel LSTM compressed-CapsNet (DC-LSTM COMP-CapsNet). Five classical classifiers are used to measure the performance of the proposed model. Outcomes display that the model of the original CapsNet exceeds the five conventional classifiers: support vector machine (SVM), K-Nearest Neighbor (KNN), radial basis function (RBF), naïve Bayes (NB), and multi-layer perceptron (MLP). Compared to our proposed model that achieves 89.3% accuracy, the highest performance is remarked by the original CapsNet with an accuracy of 84.7%.

The remainder of the paper is organized as the following: Section 2 describes prior studies and Section 3 illustrates the proposed model used for emotion recognition. Whilst the four speech databases and feature extraction utilized in this study are discussed in Section 4. Section 5 describes the outcomes obtained along with discussions. Section 6 concludes the present work.

## 2. Prior Work

### 2.1 CapsNet systems in speech applications

The CapsNet has been applied to several tasks and demonstrated its effectiveness (J. Bae & Kim, 2018a), (Zhao et al., 2018), (Turan & Erzin, 2018). Nevertheless, the CapsNet utilized in earlier work does not consider the model compression approach that is applied to CapsNet nor the implementation complexity. The used compression method in the conventional neural network cannot be directly utilized in the CapsNet. There have been also attempts to implement CapsNet to diverse areas beyond image classification, such as user intent detection ((Xia et al., 2018), (C. Zhang et al., 2019)), self-driving (Kim & Suyoung, 2019), sound event detection (Iqbal et al., 2018), speech command classification (J. Bae & Kim, 2018b), and speech emotion recognition (Wu et al., 2019a). Databases used in any neural network system should be huge and complex to arise with adequate system performance. Therefore, the work performed by Xi et al. (Xi et al., 2017) evaluated the CapsNet performance on complex data. The best validation accuracy that was reported is 71.55% trained across 50 iterations.

One of the decent applications in speech command recognition has been studied by Bae and Kim (J. Bae & Kim, 2018). The study encountered CapsNet in their system to find the spatial relationship and pose information of speech spectrogram features. The dataset that was utilized is a global single-word English dataset (Warden, 2017). The results demonstrated that the presented end-to-end speech recognition system accomplished superior results on both clean and noise-added evaluation than baseline CNN frameworks. The error rate (ER) shown in the outcomes reached down to 10.5% in a clean environment, whereas ER reached 44.7% in noisy environments. Another study by Turin and Erzin (Turan & Erzin, 2018) revealed a remarkable application of CapsNet in observing child emotional cry in local environments that can help in remote infants monitoring system. The dataset utilized is the INTERSPEECH 2018 computational paralinguistic challenge (ComParE), crying sub-challenge, which is a three-class (neutral, fussing, and crying) classification mission by utilizing an interpreted database (CRIED).

In prior work, there are four known techniques of CapsNet applications: transforming auto-encoders (G. E. Hinton et al., 2011), vector capsules based on dynamic routing (Sabour et al., 2017), matrix capsules based on expectation maximization (EM) routings (G. Hinton et al., 2018), and sequential routing framework (Lee et al., 2020). In transforming auto-encoders, any property of a picture demonstrated by a transforming auto-encoder can be manipulated. This demonstration occurs by forcing the outputs of the capsule found in a transforming auto-encoder. This idea is easily implemented, for instance, to enhance the intensity of all pixels. The vector capsules based on the dynamic routing process are explained in Section 4.1. The objective of the matrix capsules based on EM is to group several capsules to provide a part-whole relationship. The sequential routing framework is clarified in work achieved by Lee et al. (Lee et al., 2020).

Lee et al. (Lee et al., 2020) introduced the sequential routing framework (SRF), which is the first technique to adjust a CapsNet-only formation to sequence-to-sequence recognition. By presenting speech sequence recognition tasks on the TIMIT dataset, results showed an 82.6% recognition rate. This outcome is 0.8% more precise than that of CNN-based using connectionist temporal classification (CTC) networks. Wu et al. (Wu et al., 2019b) studied speech emotion recognition based on CapsNet, where the system can take into consideration the spatial association of phoneme features in spectrograms and deliver an efficient pooling technique for achieving utterance-level features. They also presented a recurrent connection to CapsNet to enhance the system time sensitivity. The evaluation has been performed on the benchmark database IEMOCAP over four emotions, i.e., sad, happy, angry, and neutral. Comparing to previous results based on combined CNN-long short-term memory (CNN-LSTM) models, results showed that the model attained better outcomes than the baseline system, where the average accuracy reached 72.73%. In another study by Tereikovska et al. (Tereikovska et al., 2019), they recognized the basic emotions by the aid of face geometry using CapsNet with an average accuracy of 85.3%.

Liu et al. (Liu et al., 2020) designed and evaluated multi-level features guided CapsNet (MLF-CapsNet) for a multi-channel electroencephalogram (EEG)-based emotion recognition. Results were tested on two databases, which are DEAP and DREAMER. DEAP is a dataset for emotion analysis physiological signals, while DREAMER represents a database for emotion recognition through EEG and electrocardiogram (ECG) signals from wireless low-cost off-the-shelf devices. In the DEAP corpus, the average reported accuracies are 98.32%, 98.31%, and 97.97% under, respectively, dominance, arousal, and valence. In the DREAMER corpus, 95.26%, 95.13%, and 94.59% are the average accuracies under arousal, dominance, and valence, respectively. GUO et al. (GUO et al., 2018) proposed a classification model based on the CapsNet neural network by obtaining the granger connection feature of original EEG signals. EEG is a sufficient modality that aids in obtaining brain signals related to several states from the scalp surface area (Kumar & Bhuvaneswari, 2012). Therefore, this study uses EEG signals as a support system to help in recognizing the correct emotion. The experimental results obtained 87.37% and 88.09% average accuracy based on valence and arousal emotion dimensions, in contrast with CNN and SVM classifier. Zhong et al. (Zhong et al., 2020) proposed a novel approach to enhance emotion recognition performance, which is called the sentiment polarity algorithm (SPT-CapsNet). Using facial expression, the accuracy of this approach reached 93.6%. The results showed that the running speed was improved, and the balance between classification accuracy and computational efficiency was maintained.

## 2.2 Classical classifiers systems in speech applications

Currently, a growing interest has been targeted to the examination of the emotional content of speech signals. Thus, several systems have been presented to identify the emotional content of the uttered sentences. The paper proposed by El Ayadi (El Ayadi et al., 2011) is a survey of speech emotion recognition focusing on three crucial facets of a speech emotion recognition model design. The first aspect is the selection of appropriate features for speech representation. The second concern is the design of a suitable classification system. The third problem is the adequate planning of an emotional speech corpus for assessing system performance. Shahin and Ba-Hutair (Shahin & Ba-Hutair, 2015) studied emotional and stressful talking condition recognition based on second-order circular suprasegmental hidden Markov models (CSPHMM2s) as a classifier. The emotional talking environment database utilized is the "emotional prosody speech and transcripts (EPST)". Results demonstrated that CSPHMM2s surpass each of the hidden Markov models (HMMs), second-order circular hidden Markov models (CHMM2s), and suprasegmental hidden Markov models (SPHMMs) with an accuracy of 76.25%.

Another work by Shahin (Shahin, 2019) proposed the utilization of "third-order circular suprasegmental hidden Markov models (CSPHMM3s)" in emotion recognition. The reported results yielded average emotion recognition accuracy of 77.8%, which was tested on the EPST dataset. Shahin (Shahin, 2016) studied the cascaded phases approach that used the speaker emotion cues endorsed by SPHMMs and HMMs as classifiers. The proposed framework contains a two-stage approach that integrates and combines the emotion recognizer complemented by a speaker recognizer into a single recognizer. The study showed that his approach gave better results with a remarkable alteration over previous studies along with other approaches such as "emotion-independent speaker verification method" and "emotion-dependent speaker verification method based on HMMs". Shahin also studied the improvement of talking condition recognition in both emotional and stressful environments based on HMMs, CHMM2s, and SPHMMs (Shahin, 2012). The attained outcomes proved that SPHMMs surpassed each of CHMM2s and HMMs using an English dataset. Besides that, outcomes also revealed that emotional talking condition recognition is much less than that in stressful talking environments by 3.3%, 2.7%, and 1.8% based on SPHMMs, HMMs, and CHMM2s, respectively.

## 2.3 Neural network systems in speech applications

Nassif *et al.* concluded in their study that prior work implemented 75% of standalone deep neural networks (DNNs) models, where only 25% of the models used hybrid models (Nassif et al., 2019). The work accomplished by Shahin *et al.* (Shahin et al., 2019) used a hybrid model, where they focused on recognizing emotion by utilizing "hybrid Gaussian mixture model and deep neural network (GMM-DNN)". GMM-DNN resulted in an accuracy of 83.97% using six different emotions, which are taken from the Arabic Emirati speech dataset. Nassif et al. (Nassif et al., 2021) aimed in their research on enhancing "text-independent speaker identification performance under both emotional and noisy talking environments". Computational Auditory Scene Analysis (CASA) and cascaded Gaussian Mixture Model – Convolutional Neural Network (GMMCNN) classifier have been employed in their study. The CASA-based module is used for the pre-processing of noise reduction, while the GMM-CNN classifier is utilized for speaker identification followed by emotion recognition. This work has achieved 83.7% accuracy using an "Arabic Emirati accented database". Mirsamadi *et al.* (Mirsamadi et al., 2017) studied automatic speech emotion recognition using recurrent neural networks (RNNs) with local attention. The intended solution assessed on the interactive emotional dyadic motion capture (IEMOCAP) corpus and proved to deliver more precise predictions compared to most of the emotion recognition systems.

## 2.4 Limitations of prior work

The limitations of similar work addressed previously were derived from a limited number of studies that focused on the application of an emotion recognition model with the employment of dual-channel LSTM compressed-CapsNet (DC-LSTM COMP-CapsNet) along with MFCCs delta-delta in harsh talking environments and the complex implementation of the model. Prior work did not solve the issues of the compression method nor the energy deficiency in emotion recognition because CapsNet structure is inefficient energy. To define the unknown emotion, an appropriate model must be constructed and discussed. Therefore, the method that is being suggested in this research to solve the previously mentioned issues poses a speech emotion recognition model based on the DC-LSTM COMP-CapsNet framework.

Most of the research works did not experience multiple real emotion states in their systems nor multiple databases for a sufficient evaluation. Moreover, the CapsNet model used in most of the prior work was not modified. There are some studies that utilized the modified CapsNet, such as the study by Zhong et al. (Zhong et al., 2020) that proposed SPT-CapsNet and the study by Liu et al. (Liu et al., 2020) that utilized MLF-CapsNet. However, SPT-CapsNet system relies on facial expression, and the MLF-CapsNet system depends on the EEG and ECG signals that are extracted from a wireless device. Our proposed system does not depend on any extra device nor on the aid of any facial expression systems.

## 2.5 The main contributions of this work

- Developing a novel DC-LSTM COMP-CapsNet classifier to ensure the model energy efficiency and adequate compression method in speech emotion recognition, which is not delivered through the original structure of a CapsNet.
- Using hyperparameter optimization based on grid search (GS) to attain optimal solutions. Results witnessed an improved performance and reduction in the training and testing running time over the trial-and-error approach.
- Creating supervised text-independent and speaker-independent DC-LSTM COMP-CapsNet model to improve the performance of emotion recognition utilizing MFCC delta-delta as feature extraction technique in abnormal talking environments.

# 3. The Proposed DC-LSTM COMP-CapsNet Framework and Emotion Recognition Algorithm

Although CNNs have a significant breakthrough in the emotion recognition field; however, CNNs need a huge training data size and are incompetent in recognizing the deformation and pose of objects guiding the establishment of capsule networks. Capsule networks are the latest phenomenon in deep learning (Punjabi et al., 2020). CapsNet has resided to this belief as its performance corresponding to the mentioned issues has surpassed CNNs (Kwabena Patrick et al., 2019).

## 3.1 The principle of capsule networks model

The concept is to incorporate "capsules" structures to CNN then re-utilize output from some of those capsules to make representations for higher capsules relatively robust (concerning different perturbations) (G. E. Hinton et al., 2011). The input to a capsule is the output features from the last CNN layer. These features are handled depending on the type of the used capsule. The output of a capsule is comprised of the probability that the feature encoded by the capsule is present and a set of vector values are commonly called "instantiation parameters". The output vector involves an observation likelihood and a position for that observation (G. E. Hinton et al., 2011). A capsule is a group of neurons that trigger independently for different forms of voice properties such as energy, pitch, etc. Technically, a group of neurons generates a capsule that is an action vector containing one part for each neuron to carry the instantiation property of that neuron (e.g., pitch) (Sabour et al., 2017). The probability of the existence of the object in a given input is the length of the vector, whereas the direction of the vector quantifies the properties of the capsule (Sabour et al., 2017), (Srihari, 2017).

Since capsules are self-sufficient, the possibility of successful identification is even greater when several capsules consent. Just once in a million trials will two capsules (a small cluster) by chance considered a six-dimensional object accepts around 10%. If the number of dimensions increases, the probability of a chance agreement exponentially decreases over a bigger cluster of higher dimensions. The outputs of capsules at lower layers are collected by higher-layer capsules and embrace those with output scalar. A cluster allows the higher capsule to create a high-level detection likelihood of an object's existence and a three-dimensional pose as well (Sabour et al., 2017). The overall block diagram is shown in Fig. 2 (Vesperini et al., 2019), where the input features activity vector is represented by a voice spectrogram. The next step remains in inserting the input vector to the convolutional layers to be inserted after that to the primary capsules. Each capsule contains an independent feature vector that has multiple neurons—one part for each neuron to carry the instantiation property of that neuron. Dynamic routing takes place next to replace the pooling step, and the last step is handled by the detection capsules.

**3.1.1 Routing by agreement**

The one capsule (child) outputs are redirected in the parent's next layer (next layer) to capsules as per the child's capability to anticipate the parent's outputs. Throughout a few simulations, the outcomes of each parent might overlap with certain children's anticipations and vary from those of others, indicating that the parent is attended or missing from the event (Sabour et al., 2017). A child evaluates a predictive vector for each potential parent by multiplying a matrix of weight (trained by backpropagation) by the parent's output (Srihari, 2017). Then, the parent's output is measured as a prediction scalar product with a coefficient reflecting the probability that this child goes to that parent.

A child whose expectations are comparatively similar to the subsequent performance raises the differential between the parent and child progressively and lowers it to balance it less well for parents. This improves the contribution the child makes to the parent. Thus, the correlation of the capsule's scalar value increases with the performance of the parent. After a small number of epochs, the coefficients firmly connect a parent to his highly probable children, suggesting that the existence of children indicates the parent's participation in the scene (Sabour et al., 2017). The more children having expectations that are similar to the performance of a parent, the more the coefficients will rise and accelerate convergence. The parent's pose (represented in its output) increasingly becomes consistent with that of its children (Srihari, 2017).

The priors, together with the weights, can be educated discriminatively. The priors rely on the parent and child capsules' position and type but not on the present data. The coefficients are modified by a "routing" Softmax for each iteration such that they sum to 1 (to represent the chance that a provided capsule is the parent of a provided child). The Softmax amplifies the bigger value and decreases lower values further than their proportion of the sum. Likewise, the likelihood of a feature is present in the input is exaggerated by a nonlinear "squash" function, which decreases values and normalizes them (dramatically smaller ones and bigger ones, but they are less than 1) (Srihari, 2017). The squash equation can be described as follows (Lee et al., 2020):

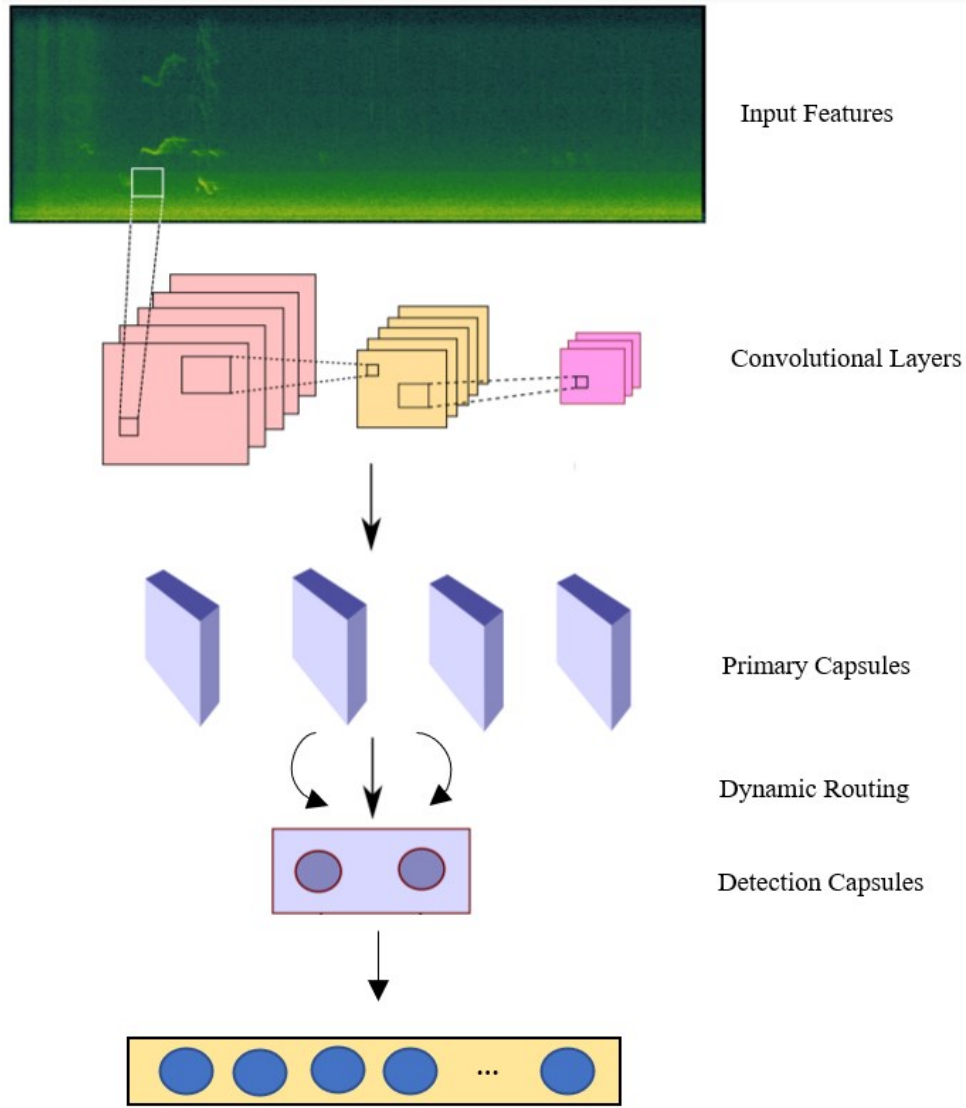

**Fig. 2** The overall architecture of CapsNet (Vesperini et al., 2019)

$$squach\,(s_j) = \frac{\|s_j\|^2}{1+\|s_j\|^2} \frac{s_j}{\|s_j\|} \quad (1)$$

where $s_j$ is the un-normalized instantiation parameter vector for the j-*th* capsule of the output capsule group. The sum of the predictions from all capsules in the lower layer is fed to capsules $s_j$ in the subsequent higher level, each with a coupling coefficient $c_{ij}$ (Sabour et al., 2017):

$$s_j = \sum c_{ij}\, \hat{u}_{j|i} \qquad (2)$$

The pose vector $u_i$ is translated and rotated by a matrix $W_{i,j}$ into a vector $\hat{u}_{j|i}$ that anticipates the parent capsule output as provided in the given equation (Sabour et al., 2017):

$$\hat{u}_{j|i} = W_{i,j}\, u_i \qquad (3)$$

### 3.2 Eliminating irrelevant instantiation parameters by compression of the model

In contrast to the traditional CNNs, the number of nodes that take part in the computation of the CapsNet is precipitously rising. The issue of expanding will turn out to be a challenging task to be ignored as the number of neural network layers is expanded. The "high-dimensional stereo" composition of the CapsNet obstructs the lower network structure of the capsule network. The capsule units of CapsNet are vector carriers, including a variety of characteristic neurons, each of which is utilized to identify a particular attribute (instantiation parameters). In addition to the strength of the target object, instantiation parameters can also signify the probability of the target object in the detection zone and the spatial position information. Scheming as various parameters as possible with hyper-parameter selection in the CapsNet leads to an incredibly conventional system; however, this design may generate excessive or irrelevant instantiation parameters.

The major concept of model compression for CapsNet is acquiring these excessive or irrelevant instantiation parameters and later "branching" these instantiation parameters. Nevertheless, by applying the CapsNet as a classifier, the number of instantiation parameters that must be in a capsule is defined by the developers that are based on application scenarios and experience. There are situations of over-design or inadequate design that can lead to insufficient results. Frequently, the insufficient design will cause an inadequate performance of the CapsNet, which is regularly altered throughout the growth phase of the network. By summarizing as various entity features as possible and design numerous instantiation parameters, the greatest potential accuracy can be obtained. Screening and compressing these additional instantiation parameters are ideas of CapsNet model compression (Zhong et al., 2020). Regarding this "overdesign" issue, this work introduces an approach like the dropout mechanism for each instantiated parameter. In the dropout mechanism, zeroing the weights of instantiation parameters is the main procedure. On the other hand, the weight will also take part in the feedback method, fine-tuning, and weight update. Therefore, throughout the network adjustment, the process is still not completed.

As an alternative to zeroing the weight of the instantiated parameter, the proposed algorithm closes the instantiation parameter itself. CNN utilizes the layer output loss function before and after modifying the weight to define if the modification has a value or the modification has to be withdrawn in order to prevent the accuracy loss, which is not appropriate. Therefore, we introduce a CapsNet framework based on the compression of instantiation parameters and energy efficiency with the optimization of the model complexity "DC-LSTM COMP-CapsNet" as described in Section 3.1.

### 3.3 The proposed DC-LSTM COMP-CapsNet algorithm

A dual-channel LSTM compressed-CapsNet (DC-LSTM COMP-CapsNet) algorithm has been proposed and implemented in our current work based on the structural features of the CapsNet to solve the previously mentioned issues. The model can compress the scale of system computation based on preserving the accuracy of system recognition, consequently, decreases the computational complexity. Dual-Channel LSTM layers intend to extract and compress sequence-correlated characteristics of Amplitude/Phase signal components and In-phase/Quadrature signal components. The distinction between our proposed novel technique and the work in (Zhong et al., 2020), is that we utilize a dual-channel LSTM approach, as well as MFCCs feature extraction is applied to compare the feature efficiency and then concatenate DC-LSTM with MFCCs output. This approach is followed by removing the outliers before inserting the feature vector into the COMP-CapsNet (compressed CapsNet). By doing so, CapsNet would result in a sufficient performance with ideal compressed input features to CapsNet.

Although nowadays the transformer techniques achieve state-of-the-art in several applications, the transformers are more effective in Natural Language Processing (NLP) problems. On the other hand, Capsule Nets that are based on dynamic routing, are more powerful in Speech Recognition (SR). The transformer techniques are the first sequence transduction models based fully on attention, substituting the recurrent layers commonly utilized in encoder-decoder architectures with multi-headed self-attention (Vaswani et al., 2017). In contrast, the idea of CapsNets is to add structures known as "capsules" to CNN, then to reuse output from various of those capsules to produce more stable (with respect to various perturbations) representations for higher capsules (Sabour et al., 2017). Our method is considered better than state-of-the-art techniques in improving the original CapsNet, where the proposed model ensures the model energy efficiency and adequate compression method in SER.

Our proposed system can be demonstrated in two phases: the preprocessing and feature extraction with the DC-LSTM phase and the recognition phase. In the first phase, all the unrecognized emotion audios (input) are pre-processed from all the noisy backgrounds then the extraction of features from these voices is accomplished using MFCCs delta-delta. The extracted features (initial features) are then labeled with (initial labels), where the extracted features out of each emotion's voice are called initial features. Each of the extracted features of the emotion's voice are then labeled in order to recognize the identity of the unknown emotion at the end of the recognition process. They are called initial as there is a final feature vector after the DC-LSTM step and after the elimination of the instantiation parameters.

Dual-channel LSTM compressed capsule network algorithm contains input and output. The input of the DC-LSTM block is the unrecognized emotion audio that comprises capsules that are included in the solid capsule layer. Each capsule has n instantiation parameters. The output of the DC-LSTM block is a list of instantiation parameters that should be eliminated. This output is inserted into the COMP-CapsNet as an input to the second phase. In the second phase, two consecutive CNN layers are designed to be inserted into the CapsNet. The output of the proposed model is the recognized unknown emotion. The proposed algorithm of DC-LSTM COMP-CapsNet is shown in Fig. 3. In the proposed model, there are pre-steps for compression and DC-LSTM features extraction before the CapsNet architecture to ensure the model energy efficiency and adequate compression method. On the other hand, the original CapsNet does not have such pre-steps.

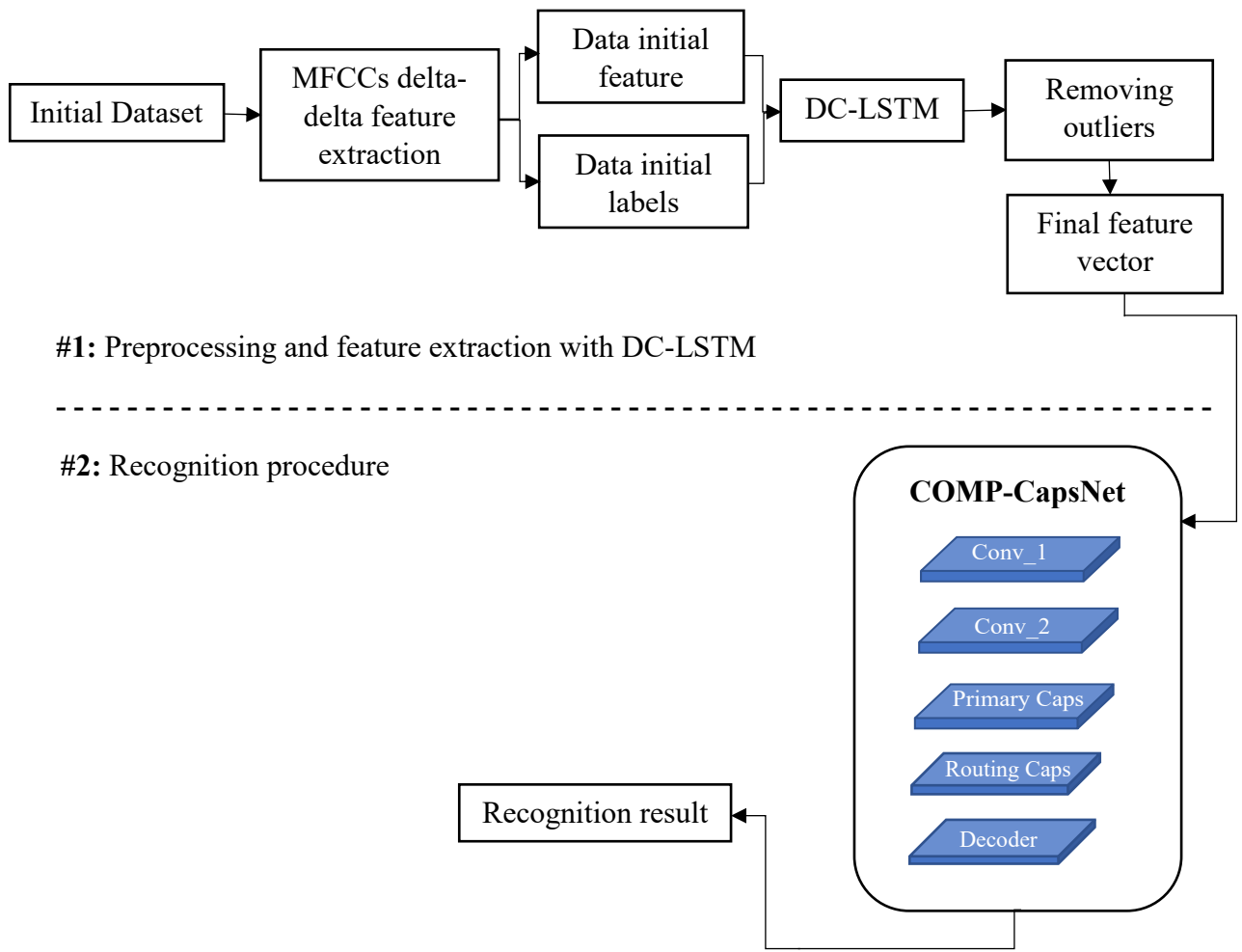

**Fig. 3** Proposed DC-LSTM COMP-CapsNet algorithm

This algorithm will assess the user in the compression method that is employed to CNN, and it helps in energy deficiency and system complexity issues by using the novel DC-LSTM COMP-CapsNet approach. Our developed pseudo-code is shown as follows:

**DC-LSTM COMP-CapsNet algorithm**

**READ** initial features data and labels

for each feature vector:

    for pitch feature in the first channel (M):

        **RETURN** LSTM produces feature batch M

    for energy feature in the second channel (N):

        **RETURN** LSTM produces feature batch N

Fusion feature of batch M and N

**SOFTMAX** layer

for all the instantiation parameters that should be eliminated from each capsule **(INPUT)**:

    for each capsule layer in the solid capsule layer:

    for variance i in an instantiation parameter vector

    for a capsule with a vector length of n:

    for the dropout matrix is set to 1 except for the value of i:

        $D = 1, D(i) = 0$

        D is multiplied with each instantiated parameter

If an instantiation parameter is trimmed out, the value of dropout is small, thus the instantiation parameter becomes meaningless (unimportant) due to its small impact.

**RETURN** all the compressed instantiated parameters based on the compression rate **(OUTPUT)**

Two consecutive convolutional layers following by primary Caps, routing Caps, and decoder

**END**

To illustrate more, first, the spoofed voices and their corresponding speaker identity are inserted into the system. After that, a pre-processing step is applied, where the dataset is divided into the training phase and testing phase, as well as organizing the voice with a label that corresponds to the speaker identity. In order to extract the important features and to reduce the complexity of the system, a feature extraction method is applied, which is a concatenation of the "MFCCs, MFCCs delta, and MFCCs delta-delta". At this point, the data is ready to be inserted into the system explained above. There is an important step between DC-LSTM and the COMP-CapsNet to remove the outliers. DC-LSTM aims to extract and compress sequence-correlated characteristics of Amplitude/Phase signal components and In-phase/Quadrature signal components. Each channel in the dual channel LSTM is comprised of two layers of LSTM each of which is made up of the same number of expanded nodes (pitch, energy) according to the input data. Corresponding to each training accuracy, the dual channel LSTM reduces the loss to attain the best biases and weights. The DC-LSTM model is divided into two channels where one channel takes energy data and the other one takes pitch data as input in batches. Each channel determines different number of LSTM cells in terms of the energy and pitch variation of the input ($M$ and $N$). By conducting synchronous training, the output feature of the final cell from the second layer of LSTM ($O_M$, $O_N$) is obtained to connect into a feature vector that can foresee each class's probability. The proposed DC-LSTM is exemplified in Fig. 4.

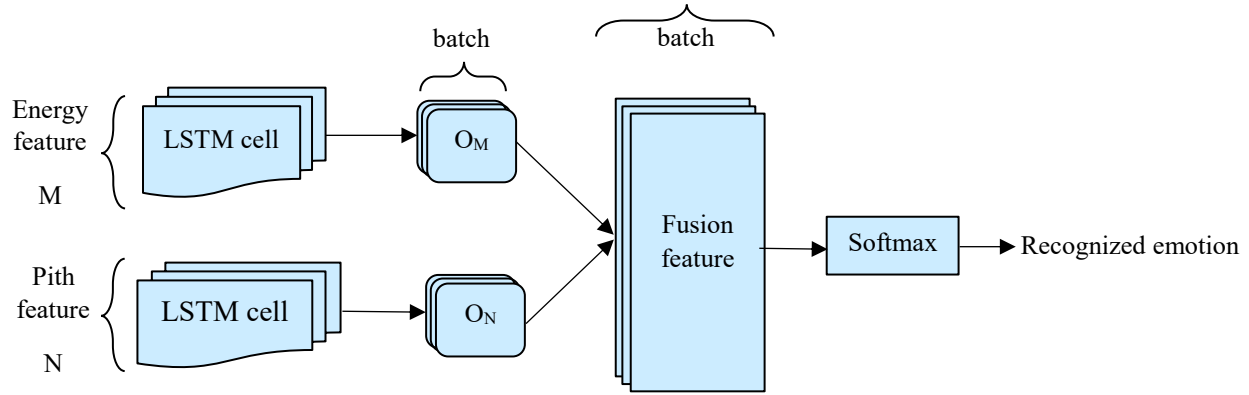

**Fig. 4** Dual channel LSTM model

### 3.4 Model parameters

Our work utilizes the powerful framework named "Keras" that is based on "Tensorflow". Based on the grid search (GS) approach that optimizes hyperparameters, we reached optimal solutions, where the number of hidden layers, input neurons, and output neurons were accurately selected. Grid search is a tuning methodology that attempts to compute the optimal values of hyperparameters in a pre-defined search grid space. It is a massive search that is functioned on all parameter values in the grid of a specific model. In GS, the domain of the hyperparameters is divided into a discrete grid. After that, by computing some performance metrics using cross-validation, every combination of values of this grid has been tried. The optimal combination of values for the hyperparameters is the goal of the grid that maximizes the average value in cross-validation. Grid search is a thorough method that traverses all the combinations to find the best point in the domain (Bergstra & Bengio, 2012). An incorrect selection of the hyperparameter values may yield incorrect results and a model with inferior performance. Therefore, by using GS algorithm in this work, our model performance has been enhanced.

The CNN model utilizes two layers. After that primary capsule with 'squash' activation also has one layer. In the third layer, the rotation algorithm capsule is occupied. In the end, a decoder network is built to achieve the final system predictions. The number of input neurons is 1,024 in the first stage and eight-dimensions capsule length in the second stage. The number of output neurons is 512 in the first output layer and 1,024 in the second output layer. After categorizing the testing labels, the output neurons represent the final classes. A normalized exponential function, "Softmax" also known as "softargmax" is an activation function that facilitates the anticipation of probability scores throughout classification tasks (Goodfellow et al., 2016). In our represented system, "Softargmax" is used and employed in the output layer. Firstly, "Softargmax" takes K real numbers (vector input), then normalizes it into a probability distribution encircling of K probabilities linked to the exponentials of the input numbers. The standard unit "Softmax" function σ: R^K→ R^K is given as the following (Wikipedia, 2014):

$$\sigma(z)_i = \frac{e^{z_i}}{\sum_{j=1}^{K} e^{z_j}} \ for\ i = 1, \dots, K\ and\ z = (z_1, \dots, z_k) \in R^K \quad (4)$$

The addition of every input vector value applied by the exponential function is then divided and normalized.

### 3.4.1 The unique parameters used in the classical classifiers

Designing the classical classifiers depends on the application that is being used. The classical classifiers contain parameters that can be altered based on the system's requirements. This section explicitly explains the parameters that have been utilized. The following are the parameters used in each classical classifier:

- **SVM**: C=1, kernel='rbf', gamma=0.001, cache_size=200, "penalty='l2', loss='squared_hinge', max_iter=1000", decision_function_shape='ovr'.

where **'C=1.0'** is the regularization parameter, where the strength of the regularization is inversely proportional to C and must be strictly positive. **'Kernel'** parameter specifies the kernel type to be used in the algorithm. It must be one of 'linear', 'poly', 'rbf', 'sigmoid', 'precomputed' or 'callable'. Moreover, **'gamma'** is the kernel coefficient for 'rbf', 'poly', and 'sigmoid'. Whereas **'cache_size'** identifies the size of the kernel cache. **'Penalty'** parameter defines the trade-off between maximizing the classification margin and minimizing the training error rate. Furthermore, the **'loss'** parameter can be indicated by the loss formula as follows:

$$loss = \max(0, predicted - original + 1) \qquad (5)$$

The better classifier model will have a lower loss. The **'max_iteration'** exemplifies the hard limit on iterations within the solver. Additionally, **'decision_function_shape'** specifies whether to return a one-vs-rest ('ovr') decision function as all other classifiers or the original one-vs-one ('ovo') decision function[1] (Lin et al., 2008).

- **MLP**: "hidden_layer_sizes=(100,), activation='relu', solver='adam', alpha=1, max_iter=200", batch_size='auto', learning_rate='constant', learning_rate_init=0.001, tol=0.001, early_stopping=True, validation_fraction=0.1, epsilon=1e-08.

where, **'hidden_layer_sizes'** is a tuple parameter, where the i[th] element signifies the number of neurons in the i[th] hidden layer. The activation function for the hidden layer is defined as **'activation='relu''**. The 'relu' is the rectified linear unit function, and it returns f(x) = max(0, x).

---

[1] https://scikit-learn.org/stable/modules/generated/sklearn.svm.SVC.html

The 'solver' is used for weight optimization. In our work, the **'adam'** solver was utilized, which refers to the stochastic gradient-based optimizer. The default solver **'adam'** perfectly operates on relatively large databases in terms of both the training time and the validation accuracy. The regularization term parameter can be obtained by the **'alpha'** parameter and the maximum number of iterations can be specified by **'max_iter'**. Moreover, **'batch_size'** is the size of minibatches for stochastic optimizers. The learning rate is set to 'constant', which is a constant learning rate given by **'learning_rate_init'** parameter that controls the step-size in updating the weights. Another important parameter is the **'tol'** parameter, which is the tolerance for optimization. If the loss or score is not improving by at least 'tol' for consecutive iterations, training ends, and the convergence is deemed to be achieved. Besides, if the validation score is not improving, **'early_stopping'** parameter will terminate training and will automatically set aside 10% of training data as validation. **'validation_fraction'** parameter is only used if early_stopping is True, where it represents the proportion of training data to set aside as validation set for early stopping. The **'epsilon'** is the value for numerical stability in the 'adam' solver[2].

- **KNN**: n_neighbors=10, weights='uniform', algorithm='auto', metric='euclidean', n_jobs=-1.

where **'n_neighbors'** is the number of neighbors to be used. A uniform weight is applied, where all points in each neighborhood are equally weighted. In order to compute the nearest neighbors, **'algorithm'** parameter is set to auto to decide the most appropriate algorithm based on the values passed to fit the model. The distance **'metric'** used for the tree is chosen to be the 'Euclidean' distance. Using **'n_jobs'** parameter defines the number of parallel jobs to run for neighbors search, where -1 means to use all processors[3]. The main advantage of the KNN method is that it is easy to interpret and does not consume time in terms of calculations (Zhai et al., 2016). The distance between all the training samples and the test sample is the Euclidean distance that can be expressed as (X. Wang et al., 2016),

$$distance(X,Y) = \sqrt{\sum_{i=1}^{N}(x_i - y_i)^2} \qquad (6)$$

where Y is the training sample and X is the testing sample.

---

[2] https://scikit-learn.org/stable/modules/generated/sklearn.neural_network.MLPClassifier.html
[3] https://scikit-learn.org/stable/modules/generated/sklearn.neighbors.KNeighborsClassifier.html

- **RBF**: C=1000.0, cache_size=200, gamma='auto', kernel='rbf', shrinking=True, tol=0.001.

where **'C=1.0'** is the regularization parameter and **'cache_size'** is the parameter that identifies the size of the kernel cache. The **'gamma'** is the kernel coefficient that is specified as 'auto', where 'auto' uses 1/number of features [4].

- **NB**: "GaussianNB, priors=None, var_smoothing=1e-09".

where NB implements the **'GaussianNB'** algorithm for classification. A framework for integrating prior distribution with observational data is delivered by the NB classifier. Assuming the prior distribution about an indeterminate event (A), which is P(A), the likelihood of gaining an experimental outcome (B), assuming that event *A* happened, is P(B│A), and the probability of detecting experimental outcome *B*, without knowing *A* has happened, is P(B). NB classifier is utilized to define the following belief about event *A* after detecting the experiment results, P(A│B) is expressed as (Karandikar et al., 2015):

$$P(A|B) = \frac{P(B|A)\,P(A)}{P(B)} \quad (7)$$

'priors' are the prior probabilities of the classes. If priors are determined, the priors are not modified according to the data. Additionally, the 'var_smoothing' parameter is the portion of the largest variance of all features that are added to variances for stability calculation [5].

### 3.4.2 The common parameters used in the classical classifiers

- (fit parameter): 'fit (X,y)' fits the model to data X and labels y, where X is an array for the training vectors of shape (n_samples, n_features). The number of samples is signified by 'n_samples' and the number of features is represented by 'n_features'. The target values are labeled in the y array.
- (score parameter): 'score (X,y)' gives back the mean accuracy on the provided test data and labels for data X and labels y.
- (predict parameter): 'Predict (X)' perform classification on samples in X to predict using the targeted classifier the labels y.

---

[4] https://scikit-learn.org/stable/auto_examples/svm/plot_rbf_parameters.html
[5] https://scikit-learn.org/stable/modules/generated/sklearn.naive_bayes.GaussianNB.html

# 4. Speech Databases and Feature Extraction

## 4.1 Arabic Emirati-accented speech corpus

The Arabic corpus utilized in this work is an Emirati-accented dialect consisting of 50 Emirati speakers (25 male and 25 female) between the age range of 20 to 55 years. Moreover, besides the neutral state, the corpus retains 5 different emotions, which are fear, happiness, sadness, disgust, and anger. Each emotion is pronounced in eight distinct utterances. Each utterance is repeated 9 times. These utterances are frequently used in the "United Arab Emirates". Skilled engineers recorded the corpus with the co-operation of "College of Communication at the University of Sharjah in the UAE". Table 1 displays the corpus utilized in this analysis, where the left column displays the Emirati accent utterances, while the right column demonstrates the corresponding English translation (Shahin, 2018b).

**Table 1**

"Arabic Emirati database and it's English translation"

| Utterance Number | "Emirati Sentences | English Translation |
|---|---|---|
| 1 | بنتلاقى ويّاك عقب ساعة | We will meet with you in an hour |
| 2 | سير عند أبويا يباك | Go to my father he wants you |
| 3 | هات تلفوني من الحجرة | Bring my cell phone from the room |
| 4 | مشغول/مشغولة الحين برمّسك عقب | I am busy now I will talk to you later |
| 5 | كل بيّاع يمدح سوقه | Every seller praises his market |
| 6 | الغريب ذيب وعضته ما تطيب | A stranger is a wolf whose bite wound will not heal |
| 7 | ناس احشمهم وناس احشم نفسك عنهم | Show respect around some people and show self-respect around other people |
| 8 | اللي ما قدرت تييبه لا تعيبه واللي ما تطوله لا تحوم حوله | Don't criticize what you can't get and don't swirl around something you can't obtain" |

## 4.2 SUSAS corpus

Hansen captured speech under simulated and actual stress (SUSAS) database, funded by the air force research laboratory (AFRL), at the University of Colorado-Boulder. The primary purpose of the SUSAS was initially to investigate the comprehension of speech in difficult contexts (J. H. L. Hansen & Bou-Ghazale, 1997). The corpus is broken down into four parts, containing a wide broad variety of stresses and emotions. Given this corpus, 32 actors, of which 19 are males and 13 are females between the age range of 22 to 76, uttering 16,000 utterances (J. Hansen, 1999).

### 4.3 RAVDESS corpus

The "Ryerson audio-visual database of emotional speech and song (RAVDESS)" collection is composed of 7,356 recorded files. The audio-only files were included in this archive, and the audio-visual files were omitted. The RAVDESS is comprised of 24 (12 males and 12 females) professional speakers between the age of 21 and 33 years, presenting two lexically matched speeches in a neutral North American accent. In RAVDESS, the speech files include 1,440 audios, while 1,012 audios are in the song files. The cumulative number of audios employed in the proposed function is thus 2,452 audios. The remaining 4,904 are removed since they are audio-visual files. This corpus includes six emotions, which are sad, angry, happy, neutral, disgusted, and fearful (Livingstone & Russo, 2018).

### 4.4 CREMA-D corpus

CREMA-D stands for crowd-sourced emotional multimodal actors dataset. This database is global English that consists of 7,442 audios from 91 speakers. These audios were collected from 48 male and 43 female speakers between the ages of 20 and 74 coming from different races (African American, Asian, Caucasian, Hispanic, and Unspecified). Speakers uttered a collection of 12 utterances. The utterances were spoken in six distinct emotions (disgust, anger, happy, sad, fear, and neutral) (Cao et al., 2014).

### 4.5 Trained/tested the model

Databases are divided into training and testing files. Therefore, there are two main phases: the training phase and the testing phase since the proposed model is text-independent and speaker-independent (the speakers used in the testing phase are different from those utilized in the training stage).

Using the Arabic Emirati-accented corpus, the training phase includes one reference model per emotion that was created using only 37 speakers out of 50 speakers producing the first 4 sentences, with nine times repetition per sentence. The overall number of utterances that have been employed in this phase is 7,992 (37 speakers × 4 utterances × 9 times/utterance × 6 emotion states). When the model was trained, the system was tested on a different number of speakers (35, 36, 37, 38, and 39), where 37 speakers were the best number that results in higher accuracy. In the test stage, each one of the 13 remaining speakers uses the second 4 utterances with a repetition of 9 times per utterance under each of the six emotions. The total number of utterances that have been utilized in this phase is 2,808 (13 speakers × 4 utterances × 9 times/utterance × 6 emotion states).

The same procedure is applied to the SUSAS database, where in the training phase, the size of the dataset is 6,600 (22 speakers × 5 utterances × 10 times/utterance × 6 emotion states). In the testing phase, the total number of utterances is 3,000 (10 speakers × 5 utterances × 10 times/utterance × 6 emotion states).

The training and testing phases in RAVDESS also follow the same procedure. "The total number of utterances in the training phase is 1,176 (14 speakers × 2 utterances × 7 times/utterance × 6 emotion states), while the total number of utterances in the testing phase is 840 (10 speakers × 2 utterances × 7 times/utterance × 6 emotion states)".

CREMA-D corpus is divided into training and testing files. "Total number of utterances used in the training phase is 2,160 (60 speakers × 6 utterances × 6 emotions), whereas the total number of utterances used in the testing phase is 1,116 (31 speakers × 6 utterances × 6 emotions)".

The training and testing phases of shallow classifiers and the proposed classifier were achieved. "In the training phase, the proposed classifier and the shallow classifiers employ the features of training files to fit them with their targets. Whereas in the testing phase, the features of testing files are utilized. The features of testing files are used to fit them with their targets".

## 4.6 Feature extraction

Linear predictive coding (LPC) (O'Shaughnessy, 1988), (Dave, 2013), Mel-Frequency Cepstral Coefficients delta-delta (MFCCs delta-delta) (Davis & Mermelstein, 1980), Hybrid Algorithm DWPD (Sunny et al., 2013), discrete cosine transforms (DCT) (Sahidullah & Saha, 2009), and probabilistic linear discriminate analysis (PLDA) (Ioffe, 2006), (Narang & Gupta, 2015) are the five techniques of feature extraction that are used to determine the best output of dimensionality in our work. Neither approach surpasses the other; however, the best approach should be chosen based on the result of the model performance. MFCC acknowledges the best output feature in the proposed work, where various enhancements to the MFCC method were introduced to make them less sensitive to noise, more stable, and quicker. Further details are provided in Experiment 9 of Section 5.

The extraction of features in this study utilizes a concatenation of "delta-delta, MFCCs-delta, and MFCCs". The MFCC is a representation of an audio signal short-term power spectrogram function (with other tune-able parameters defined for nested core functions), all parameters related to wave signal segmentation (such as the overlapped values and frames) are mainly used.

To choose functions, the command is: "compute-mfcc-feats. Therefore, the MFCC system was utilized for selecting functions. The output is, by default, 20-dimensional selected features. The 20-dimensional is inadequate to represent a very broad set of data, the dimensionality to 40-dimensional chosen features has been expanded. MFCCs extracted feature are fine, but other MFCCs can work more skillfully, such as adding coefficients of dynamic features such as delta-delta and delta (first order and second order frame-to-frame difference). Typically, an improvement in efficiency or output performance levels of 20.0% arises when MFCC delta-delta features are utilized relative to MFCC features as they have a wealthier frame context (Ahmad et al., 2015), (Godino-llorente et al., 2006). A concatenation of MFCC, MFCC delta, and MFCC delta-delta were utilized in the proposed work".

The rate of sampling is Sr, "the log-power Mel spectrogram is S, n_mfcc is the number of MFCC in return, and y denotes the wave file. There are n_fft=512 samples (physical duration: 23 ms at sr= 22,050 Hz sample rate), and hop length= 256 samples to get 50% overlap for each frame. By default, the Dct_type is the discrete cosine transform (DCT), and the DCT type-2 is usually used. If dct_type is 2 or 3, making norm='ortho' uses an ortho-normal DCT basis. **kwargs is utilized for further keyword arguments. MFCCs-delta-delta holds richer background information that contributes to better precision than each of MFCC and MFCC-delta (Hanson & Applebaum, 1990), (Furui, 1986). MFCC-delta-delta has a lower number of coefficients than each of MFCC and MFCC-delta".

## 5. Results and Discussions

Our novel work proposes a DC-LSTM COMP-CapsNet framework that enhances text-independent and speaker-independent emotion recognition performance under six distinct emotions using four diverse databases. The emotional talking conditions are: happy, sad, disgust, angry, fearful, and neutral. Nine numerous experiments are conducted through this study: average emotion recognition performance based on different classifiers, ROC curve, precision, recall, confusion matrix, statistical test, comparisons between the proposed framework and each of the prior work, CNN model alone, the original CapsNet model with different datasets, and computational complexity. Our proposed classifier is assessed using: SUSAS, RAVDESS, and CREMA-D corpora, human listener, and diverse feature extraction approaches.

1) **Experiment 1: Emotion recognition model assessment based on various classifiers**

A) **Model accuracy**

Table 3 validates average emotion recognition performance based on various classifiers using the Arabic Emirati-accented corpus. The highest accuracy amongst the classical classifiers is reported for SVM, where the accuracy reaches 69.8%, followed by MLP that achieves 69.2%. On the other hand, RBF and NB remark the lowest accuracy. Nevertheless, the proposed framework attains an accuracy of 89.3%, which shows that the proposed work is superior to each of the classical classifiers, CNN and CapsNet.

It is well known that classifiers work best in neutral talking conditions than in any harsh talking environment such as emotional talking environments. Fig. 5 shows that the greatest performance occurs in a neutral state and oppositely in the rest of the talking states using the Emirati-accented corpus as listed in Table 2. The average of each classifier has been computed, and results indicate that the proposed model proves its superiority based on each of CNN, CapsNet, MLP, KNN, NB, RBF, and SVM in all talking conditions. Consequently, our proposed architecture reports the optimum performance, to date, due to its capability to adequately accomplish the compression step and simultaneously maintain sufficient energy with low complexity. Furthermore, the results of emotion recognition performance using the Emirati-accented database demonstrate that the proposed framework has homogenous (close performances) outcomes across different emotional talking conditions, whereas non-homogenous results appear for the other classifiers.

**B) ROC curve**

A substantial comparison among shallow classifiers, CNN, CapsNet, and DC-LSTM COMP-CapsNet model using the ROC (receiver operating characteristic) curve of the traditional classifiers is demonstrated in Fig. 6. ROC curve signifies the graphical structure that illustrates the analytical ability of a classifier composition as its discernment threshold is altered. As shown in Fig. 6, SVM has the maximum area under the curve, followed by the MLP, whereas RBF has the least accuracy as it has the minimum area under the curve.

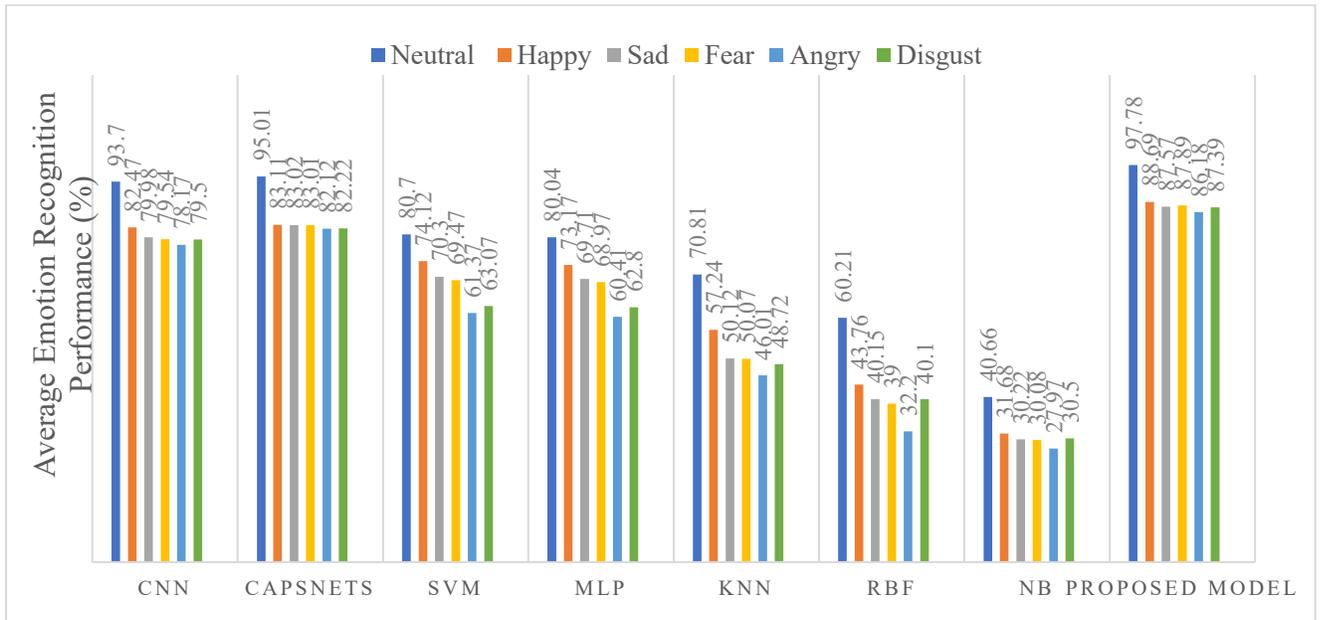

**Fig. 5** Emotion recognition performance assessment using Arabic Emirati-accented corpus based on CNN, CapsNet, SVM, NB, RBF, MLP, and KNN

**Table 2**
Average emotion recognition accuracy using Arabic Emirati-accented corpus based on diverse classifiers

| Classifiers | Performance (%) |
|---|---|
| MLP | 69.2 |
| KNN | 53.8 |
| SVM | 69.8 |
| RBF | 42.6 |
| NB | 31.9 |
| CNN | 82.2 |
| CapsNet | 84.7 |
| Proposed Model | 89.3 |

## C) Precision, recall, and confusion matrix

Precision (correctness) and recall (completeness) are two essential parameters in defining the performance of the proposed system. The recall is the number of related records recovered by an examination divided by the total number of current related records, whereas precision is the number of related records recovered by an examination divided by the total number of records recovered by that examination (Buckland & Gey, 1994). Precision is the portion given by correctly predicted positive clarifications to the predicted positive interpretations, meanwhile, recall is the portion of truly predicted positive clarifications to the overall negative interpretations as demonstrated in Fig. 7. Precision and recall functions are, respectively, demonstrated as the following (Goutte & Gaussier, 2005):

$$precision = \frac{TP}{TP+FP} \qquad (8)$$

$$Recall = \frac{TP}{TP+FN} \qquad (9)$$

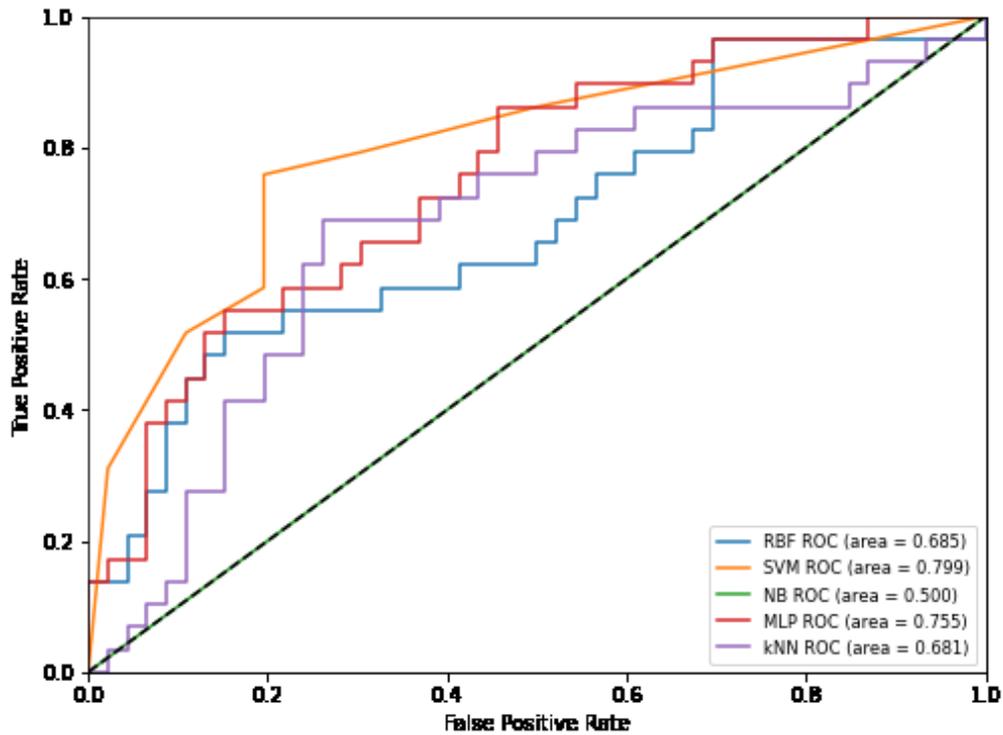

**Fig. 6** ROC curve representation of the five classical classifiers

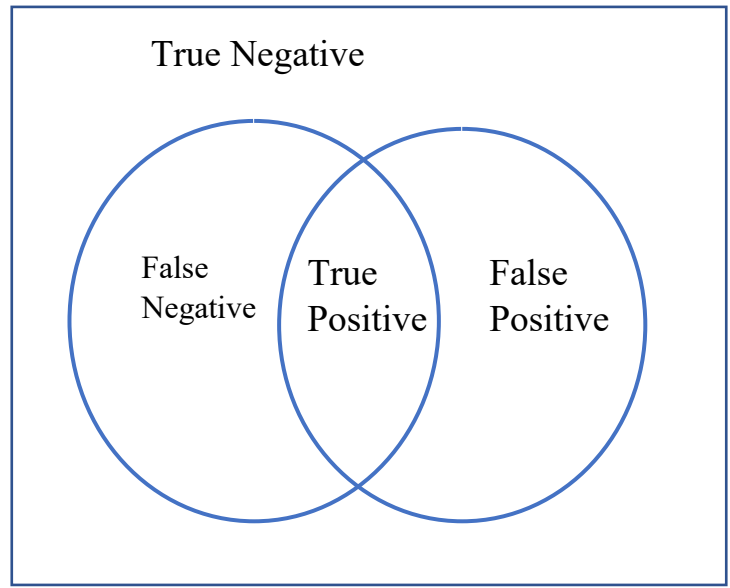

**Fig. 7** Precision and recall

Given, *TP* (true positive) resembles the truly predicted positive values. This indicates that the values of the real class are correct, as well as for the value of the predicted class. On the other hand, *TN* (true negative) states the truly predicted negative values. This indicates that the value of the real class is incorrect, as well as for the value of the predicted class. *FP* (false positive) denotes when the real class is incorrect, but the predicted class is correct. On the contrary, *FN* (false negative) where the real class is incorrect, but the predicted class is incorrect (Goutte & Gaussier, 2005).

The outcomes of recall for the proposed framework, the original CapsNet, CNN, and the shallow classifiers using the Emirati-accented database, are given in Table 3. Furthermore, the outcomes of precision for the proposed system, CapsNet, CNN, and the shallow classifiers are demonstrated in Table 4 utilizing the same dataset. The findings in Table 3 reveal that the maximum recall applies to the proposed model, CapsNet, CNN, SVM, MLP, KNN, RBF, and NB, respectively. In comparison, recall outcomes do well in neutral, happy, sad, fear, disgust, and angry state, respectively.

**Table 3**

Computed recall for shallow classifiers, CNN classifier, CapsNet model, and the proposed model in the six emotional states using Emirati-accented corpus

| Model/State | Neutral | Fear | Sad | Happy | Disgust | Angry | Average |
|---|---|---|---|---|---|---|---|
| **Proposed model** | 0.99 | 0.76 | 0.77 | 0.80 | 0.77 | 0.71 | 0.80 |
| CapsNet | 0.95 | 0.75 | 0.76 | 0.77 | 0.75 | 0.70 | 0.78 |
| CNN | 0.90 | 0.74 | 0.76 | 0.76 | 0.73 | 0.68 | 0.76 |
| SVM | 0.85 | 0.68 | 0.70 | 0.70 | 0.65 | 0.66 | 0.71 |
| MLP | 0.83 | 0.61 | 0.60 | 0.71 | 0.56 | 0.53 | 0.64 |
| KNN | 0.83 | 0.46 | 0.48 | 0.70 | 0.45 | 0.42 | 0.56 |
| RBF | 0.59 | 0.31 | 0.33 | 0.35 | 0.30 | 0.27 | 0.36 |
| NB | 0.47 | 0.25 | 0.26 | 0.28 | 0.24 | 0.20 | 0.28 |

**Table 4**

Computed precision for shallow classifiers, CNN classifier, CapsNet model, and the proposed model in the six emotional states using Emirati-accented corpus

| Model/State | Neutral | Fear | Sad | Happy | Disgust | Angry | Average |
|---|---|---|---|---|---|---|---|
| **Proposed model** | 0.98 | 0.76 | 0.78 | 0.85 | 0.76 | 0.74 | 0.81 |
| CapsNet | 0.92 | 0.73 | 0.77 | 0.83 | 0.74 | 0.73 | 0.79 |
| CNN | 0.89 | 0.70 | 0.79 | 0.81 | 0.72 | 0.71 | 0.77 |
| SVM | 0.71 | 0.51 | 0.52 | 0.57 | 0.50 | 0.48 | 0.55 |
| MLP | 0.72 | 0.53 | 0.54 | 0.55 | 0.49 | 0.47 | 0.55 |
| KNN | 0.63 | 0.51 | 0.52 | 0.54 | 0.50 | 0.41 | 0.52 |
| RBF | 0.45 | 0.31 | 0.33 | 0.35 | 0.34 | 0.27 | 0.34 |
| NB | 0.37 | 0.25 | 0.26 | 0.28 | 0.23 | 0.21 | 0.27 |

Among all the conventional classifiers, CapsNet has achieved the highest precision in all states, as seen in Table 4. SVM has the highest precision of the shallow classifiers, while NB has the lowest. All classifiers reach the highest precision in the neutral state competed with the other states.

A confusion matrix that capitulates a confusion percentage of a test emotion with the other emotions using the Arabic Emirati-accented dataset based on the proposed model is given in Table 5. This table determines the following:

(a) Neutral emotion is the easiest recognizable emotion among other emotions (97.8%). This is in agreement with the prior work (Shahin & Ba-Hutair, 2015), (Shahin et al., 2018), (Shahin, 2012). Therefore, the highest performance of emotional talking condition recognition is neutral.

(b) Angry emotion is the least easily recognizable emotional talking condition (86.2%). Accordingly, the least talking state recognition performance in such a talking environment is angry. This result agrees with that in previous studies (Shahin & Ba-Hutair, 2015), (Shahin et al., 2018), (Shahin, 2012).

(c) In the fear emotion column, e.g., 87.9% (in bold) of the sentences that were uttered in a fear state were recognized correctly. 0% represents that the neutral state is the least state identified as fear. Therefore, fear emotion is not confusable at all with neutral emotion. This column displays that the fear state has the highest confusion percentage with the angry state (5%). Therefore, fear emotion is highly confusable with angry emotion. This column also exemplifies that 2% of the sentences that were generated in a fear emotion were recognized as disgust emotion.

**Table 5**
Confusion matrix in emotional talking conditions using the Arabic Emirati-accented database based on the proposed model (%)

| Emotion | Neutral | Happy | Sad | Fear | Disgust | Angry |
|---|---|---|---|---|---|---|
| Neutral | **97.8** | 1.5 | 1 | 0 | 0 | 1 |
| Happy | 1.2 | **88.7** | 2.2 | 2.1 | 2.5 | 1.3 |
| Sad | 0 | 1 | **87.6** | 3 | 2 | 2 |
| Fear | 0.4 | 2.3 | 4.5 | **87.9** | 1.1 | 3.1 |
| Disgust | 0.6 | 2 | 1.4 | 2 | **87.4** | 6.4 |
| Angry | 0 | 4.5 | 3.3 | 5 | 7 | **86.2** |

**D) *t*-test (statistical distribution)**

The inferential statistic is used to identify whether there is a significant variance between the means of two groups signified by the "*t*-statistical distribution test" or not. Testing of the hypothesis is considered as the major utilization of the *t*-statistical test (Investopedia, n.d.). Consequently, the t-test is conducted to verify if performance variances of emotion recognition are true or only due to statistical variabilities. The t-test is carried utilizing the Emirati-accented database to evaluate our proposed model, CapsNet, CNN, and the classical classifiers. The equation below yields the t-test (Hogg et al., 2005),

$$t_{1,2} = \frac{\bar{x}_1 - \bar{x}_2}{SD_{pooled}} \tag{10}$$

Given $\bar{x}_1$, $\bar{x}_2$, as the mean of the 1st and 2nd sample, respectively, where both have the same size '*n*'. The pooled standard deviation ($SD_{pooled}$) of the two samples is given as (Hogg et al., 2005),

$$SD_{pooled} = \sqrt{\frac{SD_1^2 + SD_2^2}{2}} \tag{11}$$

Given the standard deviation of the first sample ($SD_1$) of size '$n$', and the standard deviation of the second sample ($SD_2$) of equal size '$n$' (Hogg et al., 2005).

Table 6 displays the measured t-values of the proposed framework with each of CapsNet and CNN ($t_{proposed, CNN}$ and $t_{proposed, CapsNet}$) using the Emirati-accented corpus. It is unnecessary to quantify the *t* values between the proposed model and the remainder of the shallow classifiers (KNN, SVM, MLP, RBF, and NB), provided that SVM and MLP are the top two classifiers among the classical classifiers with the two highest results. Given this table, each computed *t* value ($t_{1,2}$) is greater than the "tabulated critical value $t_{0.01}$=3.17 at 0.01 significant level". Each measured *t*-value is larger than $t_{0.01}$=3.17, so the assessment confirms that the proposed framework generates a substantial performance enhancement for emotion recognition.

**Table 6**
Computed *t* values between the proposed model and each of CapsNet and CNN using Emirati-accented database

| $t_{1,2}$ | Calculated $t$ values |
|---|---|
| $t_{proposed, CNN}$ | 3.21 |
| $t_{proposed, CapsNet}$ | 3.19 |

**2) Experiment 2: Comparisons with different approaches**

In this experiment, two comparisons are conducted. The first comparison is between the proposed work and prior work, whereas the second one is between the proposed work and each of CNN and CapsNet.

**A) Comparison between the proposed work and prior work**

Emotion recognition results based on the proposed framework are shown in Table 7 along with those based on distinct classifiers employed in preceding studies using the Arabic Emirati-accented database. Table 8 illustrates the comparison between the proposed work and prior work using the English database. It is noticeable that the proposed model significantly surpasses other classifiers and models utilized in prior work for emotion recognition. The average relative improvement rate of the proposed model over the prior work is shown as follows:

$$\text{Average relative improvement rate} = \frac{Proposed\ work\ accuracy - Related\ work\ accuracy}{Related\ work\ accuracy} \times 100 \quad (12)$$

As demonstrated in Table 8, our results are significantly better than those reported in prior work. The study by Wu et al. (Wu et al., 2019) used CapsNets, which showed that the average improvement rate between our proposed system and their work is 22.78% which proves the superiority of our model over their model.

**Table 7**
Comparison between the proposed work based on DC-LSTM COMP-CapsNet and that based on "state-of-art" approaches in prior work of emotion recognition performance using Arabic Emirati-accented corpus

| Prior work | Classifier | Emotion recognition accuracy (%) | *Average relative improvement rate of the proposed model over prior work (%) |
|---|---|---|---|
| Shahin et al. (Shahin et al., 2019) | GMM-DNN | 83.97 | 6.35 |
| Shahin (Shahin, 2019) | CSPHMM3s | 77.80 | 14.78 |

**Table 8**
Comparison between the proposed work based on DC-LSTM COMP-CapsNet and that based on "state-of-art" approaches in prior work of emotion recognition performance using English corpus

| Prior work | Classifier | Emotion recognition accuracy (%) | *Average relative improvement rate of the proposed model over prior work (%) |
|---|---|---|---|
| Wu et al. (Wu et al., 2019) | CapsNet | 72.73 | 22.78 |
| Xi et al. (Xi et al., 2017) | CapsNet | 71.55 | 24.81 |
| Shahin and Ba-Hutair (Shahin & Ba-Hutair, 2015) | CSPHMM2s | 76.25 | 17.11 |
| Shahin (Shahin, 2012) | HMMs | 62.08 | 43.85 |
| | CHMM2s | 66.92 | 33.44 |
| | SPHMMs | 72.75 | 22.75 |

**B) Comparison between the proposed work and each of CNN and the original CapsNet**

In this experiment, the comparison between the proposed model and each of CNN and the original CapsNet will be discussed based on their performance, average running time, and code complexity. It is shown in Tables 9, 10, and 11 that the average running time of CNN and original CapsNet is approximately similar to the proposed work; however, the code complexity of the proposed model is higher than each of CNN and the original CapsNet. On the other hand, the proposed classifier reports a superior performance to each of those based on CNN and the original CapsNet. The comparison has been done on the same PC with the following specifications: "Intel® Core™ i7-9750H with a CPU @2.60GHZ (12 CPUs) ~2.6GHZ".

Table 9
Comparison between the proposed work and each of CNN and CapsNet based on different attributes using Arabic Emirati-accented corpus

| Classifier | Emotion recognition accuracy (%) | Average running time (sec) | Code complexity |
|---|---|---|---|
| CNN | 82.2 | 2.31 | Less complex |
| CapsNet | 84.7 | 2.37 | Less complex |
| DC-LSTM COMP-CapsNet | 89.3 | 2.39 | More complex |

Table 10
Comparison between the proposed work and each of CNN and CapsNet based on distinct features using SUSAS corpus

| Classifier | Emotion recognition accuracy (%) | Average running time (sec) | Code complexity |
|---|---|---|---|
| CNN | 79.5 | 1.84 | Less complex |
| CapsNet | 80.8 | 1.89 | Less complex |
| DC-LSTM COMP-CapsNet | 82.9 | 1.91 | More complex |

Table 11
Comparison between the proposed work and each of CNN and CapsNet based on diverse properties using RAVDESS corpus

| Classifier | Emotion recognition accuracy (%) | Average running time (sec) | Code complexity |
|---|---|---|---|
| CNN | 75.3 | 1.75 | Less complex |
| CapsNet | 77.9 | 1.80 | Less complex |
| DC-LSTM COMP-CapsNet | 82.1 | 1.83 | More complex |

## 3) Experiment 3: Computational complexity

The measurement of the space and/or time needed (speed) by an algorithm for an input of a provided size (n), is the complexity of an algorithm. Complexity is dependent upon factors including the computer architecture, e.g., the hardware platform representation of the abstract data type (ADT) compiler efficiency and the complexity of the underlying algorithm size of the input. ADT is a type with related operations but with hidden representations (Kastens & Waite, 1991). Built-in primitive types are the most common types of abstract data in python language, float, and integer. While the most important considerations, as will be seen later, are the complexity of the underlying algorithm and the input size. The algorithm used to calculate the computational complexity is known as the "Quicksort algorithm".

Quicksort is a sort of contrast, which means that it can sort objects of any type for which a relationship "less-than" is specified (formally, a complete order). Efficient Quicksort applications are not a stable sort, which means that the relative order of objects of the equivalent sort is not maintained (Cederman & Tsigas, 2008). The widely utilized notation to measure the algorithm running time complexity is shown below:

An algorithm complexity or outcome is calculated by the utilization of the big O notation (S. Bae, 2019). Namely, it provides information to express the upper limit of an algorithm which provides measures for the maximum possible time or worst complexity time to complete an algorithm:

- Big $\Omega$ notation, is the systematic way of representing the lower limit of the running time of an algorithm. It calculates the absolute best complexity or the best (lowest) duration of an algorithm.
- Huge notation the systematic way to represent the lower limit as well as the upper limit of the running time of an algorithm.

The Big $O$ notation is often utilized to determine the upper limit of an algorithm, while the Big $\theta$ notation is utilized at times to identify the average case and $\Omega$ notation is the least common notation among the three. After analyzing the code, the time complexity of the proposed model and all the other classifiers are listed in Table 12, where *n* is the number of emotions.

**Table 12**

Time complexity according to Big *O* notation for all the presented classifiers using the Arabic Emirati-accented database

| Models/ Case | Proposed model | CapsNet | CNN | SVM | MLP | KNN | RBF | NB |
|---|---|---|---|---|---|---|---|---|
| Best case | $\Omega(n \log(n))$ | $\Omega(n \log(n))$ | $\Omega(\log(n))$ | $\Omega(\log(n))$ | $\Omega(\log(n))$ | $\Omega(\log(n))$ | $\Omega(m^2)$ | $\Omega(\log(n))$ |
| Average case | $\Theta(n \log(n))$ | $\Theta(n \log(n))$ | $\Theta(\log(n))$ | $\Theta(\log(n))$ | $\Theta(n \log(n))$ | $\Theta(\log(n))$ | $\Theta(m^2)$ | $\Theta(\log(n))$ |
| Worst case | $O(n^2)$ | $O(n^2)$ | $O(n)$ | $O(n)$ | $O(n)$ | $O(n)$ | $O(m)$ | $O(n)$ |

4) **Experiment 4: Training and testing running time comparison**

This work also accounts for an essential aspect called running time to test the proposed model against each of the CapsNet model, CNN classifier, and the classical classifiers. The running time engaged in testing/train the mentioned models using the Emirati-accented database is shown in Table 13. As mentioned earlier, the same PC has been used with the following specifications for all the experiments: "Intel® Core™ i7-9750H with a CPU @2.60GHZ (12 CPUs) ~2.6GHZ".

Table 13 presents that the running time of the proposed model is roughly the same as all other models except for RBF and NB classifiers. Although the proposed model is not the fastest in the training phase, it is the fastest in the testing phase, which is more critical since training can be performed offline, but testing is performed online.

**Table 13**

Running time associated with each of the proposed work, CapsNet, CNN, and all the classical classifiers using Emirati-accented database

| Model | Training time (sec.) | Testing time (sec.) |
|---|---|---|
| DC-LSTM COMP-CapsNet | 2.8 | 1.8 |
| CapsNet | 2.3 | 1.6 |
| CNN | 2.2 | 1.5 |
| SVM | 1.9 | 1.3 |
| MLP | 1.9 | 1.3 |
| KNN | 1.8 | 1.2 |
| RBF | 1.7 | 1.1 |
| NB | 1.6 | 1.0 |

## 5) Experiment 5: SUSAS corpus

The studies associated with speech recognition are put in an application by the SUSAS corpus, standing for "speech under stimulated and actual stress", which is an English database known globally. The SUSAS corpus has been utilized in terms of analyzing and evaluating the proposed model, CapsNet, CNN, and the classical classifiers. Other studies employ SUSAS in order to test multiple datasets as given in Shahin's illustrated work, instead of just employing one dataset in the assessment (Shahin, 2018a).

Angry, disgust, fear, happy, sad, and neutral talking states are the six states that have been chosen in this work out of other more emotions in the SUSAS database. Average emotion recognition performance based on the proposed framework, CapsNet, CNN, MLP, KNN, SVM, RBF, and NB utilizing the six talking conditions is displayed in Fig. 8. The proposed model records the highest performance amongst all the classical classifiers, CNN and CapsNet, which are 75%, 78%, 80%, 82%, 82%, and 96% in angry, disgust, fear, sadness, happiness, and neutral emotions, respectively. The proposed model surpasses current classifiers as a result of the ability of DC-LSTM COMP-CapsNet to automatically perceive the outstanding features from raw data without any human command; thus, the proposed model will learn the features of each capsule on its own. Moreover, due to the ability of the model to manage massive and formless data, DC-LSTM COMP-CapsNet aims to revolutionize the industry and surpasses other models. The minimal performance, on the other hand, is reported by the NB classifier, at which performance is 52.7% in the neutral talking condition and 32.0% utilizing the SUSAS dataset in the angry talking condition. It can be concluded that emotion recognition accuracies using our proposed model under all emotions except for the neutral state are homogenous.

The confusion matrix that represents a confusion percentage of a test emotion state with the other emotions using the SUSAS dataset based on the proposed model is represented in Table 14. This table reveals the following:
 (a) Neutral emotion is recorded as the most simply recognizable emotional talking condition among other emotions (96.0 %). Accordingly, the highest emotion recognition performance takes place under neutral.

(b) Angry emotion is testified to be the least recognizable emotional talking condition (75.0%). This proves that the least emotion that has been recognized is angry with the lowest recognition performance.

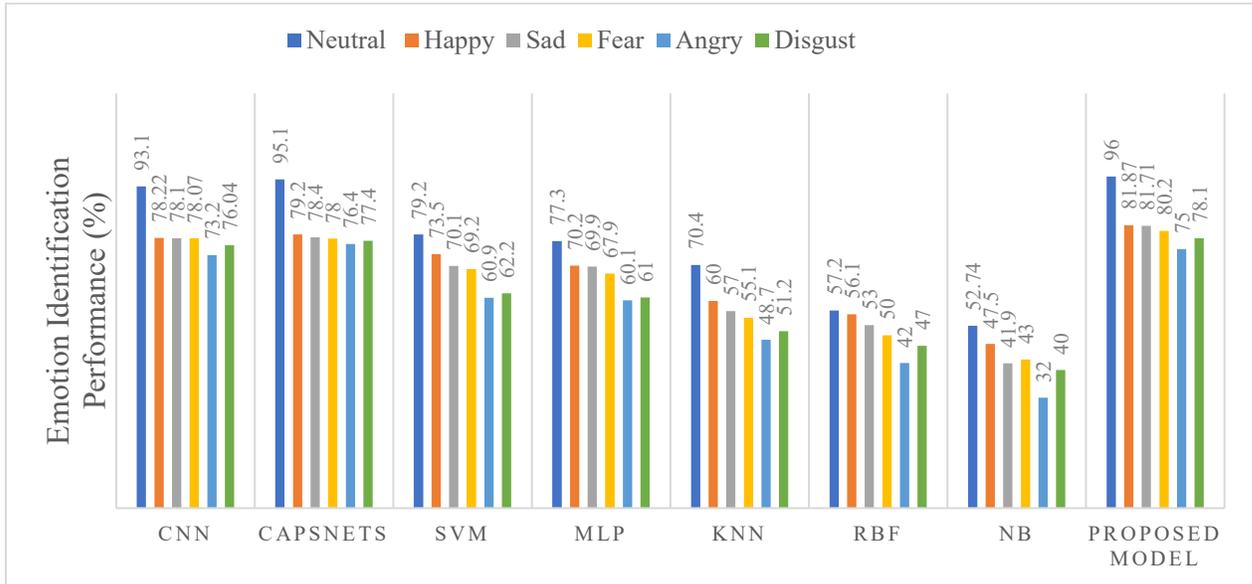

**Fig. 8** Average emotion recognition performance based on DC-LSTM COMP-CapsNet framework, CapsNet, CNN, SVM, KNN, MLP, RBF, and NB using SUSAS corpus

Table 14
Confusion matrix using the SUSAS database based on the proposed model (%)

| Emotion | Neutral | Happy | Sad | Fear | Disgust | Angry |
|---|---|---|---|---|---|---|
| Neutral | **96.0** | 8 | 1.3 | 1 | 2 | 3.5 |
| Happy | 1.5 | **81.9** | 1.5 | 2 | 1.5 | 5 |
| Sad | 1 | 2.5 | **81.7** | 6.5 | 6 | 6.5 |
| Fear | 0.5 | 2 | 7 | **80.2** | 8 | 4 |
| Disgust | 1 | 2.5 | 3.5 | 4.3 | **78.1** | 6 |
| Angry | 0 | 3.1 | 5 | 6 | 4.4 | **75.0** |

6) **Experiment 6: RAVDESS corpus**

RAVDESS points to the Ryerson Audio-Visual Emotional Expression and Song Database as an English corpus present worldwide. RAVDESS was analyzed to assess the efficiency of the proposed system for emotion recognition. A total of 2,452 audios, including the song and speech files are covered by the RAVDESS dataset. Average emotion recognition performance based on the proposed architecture, CapsNet, CNN, RBF, MLP, KNN, SVM, and NB using RAVDESS is exemplified in Fig. 8. This figure indicates that the proposed model is the best for all specified classifiers, where the efficiency shot up 95.8% in the neutral speaking condition. Among other emotions used, angry emotion has the minimum performance based on all the classifiers listed. It is possible to report an observation from Fig. 9 that CNN and CapsNet classifiers noted the next ideal performance following the proposed model in a neutral state, with 94.3% and 95.0%, respectively. In comparison, in any spoken state, NB and RBF classifiers do not report efficient results, where their ability to recognize the right emotion is thus deceivingly negligible. Emotion recognition accuracies reported by our proposed work are homogenous except for neutral and angry states.

A representation of a confusion percentage of a test emotion with the other emotions is measured by a confusion matrix based on the proposed model using RAVDESS corpus is presented in Table 15. This table exposes the following:

(a) The simplest recognizable emotion state is neutral (95.8%); thus, neutral emotion has the highest recognition performance in emotional talking environments.

(b) In contrast, angry emotion is testified as the least recognizable emotion (75.0%). This shows that the worst emotion recognition performance goes for the angry talking condition.

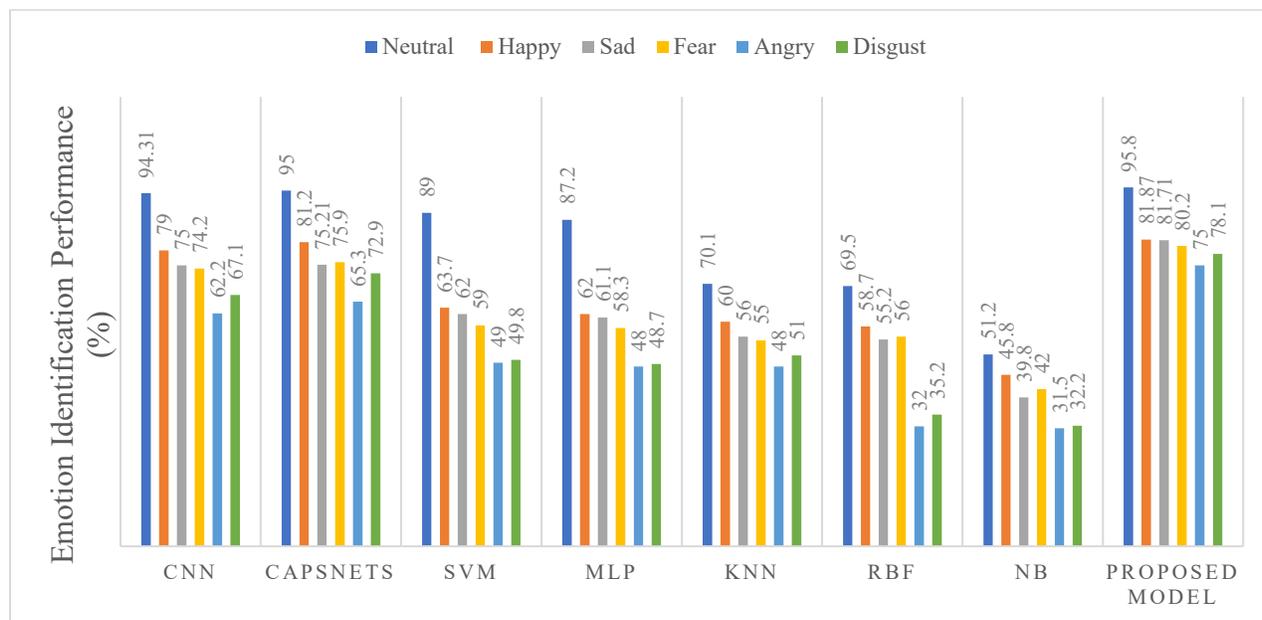

**Fig. 9** Average emotion recognition performance based on the proposed model, CapsNet, CNN, KNN, MLP, RBF, SVM, and NB using RAVDESS corpus

**Table 15**

Confusion matrix of emotional talking conditions using the RAVDESS database based on the proposed model (%)

| Emotion | Neutral | Happy | Sad | Fear | Disgust | Angry |
|---|---|---|---|---|---|---|
| Neutral | **95.8** | 8 | 1 | 2 | 2 | 3 |
| Happy | 1.5 | **81.9** | 2.3 | 4.3 | 1.4 | 4.5 |
| Sad | 1 | 1.5 | **81.7** | 6 | 6 | 7 |
| Fear | 0 | 2.5 | 6.5 | **80.2** | 8 | 2.5 |
| Disgust | 0.5 | 2 | 3 | 3 | **78.1** | 8 |
| Angry | 1.2 | 4.1 | 5.5 | 4.5 | 4.5 | **75.0** |

7) **Experiment 7: CREMA-D corpus**

Crowd-sourced emotional multimodal actors dataset (CREMA-D) is a global English database associated with speech recognition applications. This dataset has been employed in order to analyze and evaluate the proposed model, CapsNet, CNN, and the classical classifiers.

All of the six emotional states of the CREMA-D dataset (angry, disgust, fear, happy, sad, and neutral) have been utilized in this study. Average emotion recognition performance based on the proposed framework, CapsNet, CNN, MLP, KNN, SVM, RBF, and NB using the six emotions is shown in Fig. 10. The proposed model remarks the highest performance amongst all the classical classifiers, CNN and CapsNet, which are 75%, 77%, 78%, 80%, 83%, and 96% for angry, disgust, fear, sad, happy, and neutral emotion, respectively. On the other hand, the lowest performance is reported by the NB classifier, at which the performance is 53% for the neutral talking condition and 30% for the angry talking condition.

The confusion matrix based on the proposed model using CREMA-D corpus is presented in Table 16. This table reveals the following:

(a) The simplest recognizable emotion state is neutral (95.6%); thus, neutral emotion has the highest recognition performance among the remaining emotional states.
(b) In contrast, angry emotion is testified as to the least recognizable emotion (75.6%).

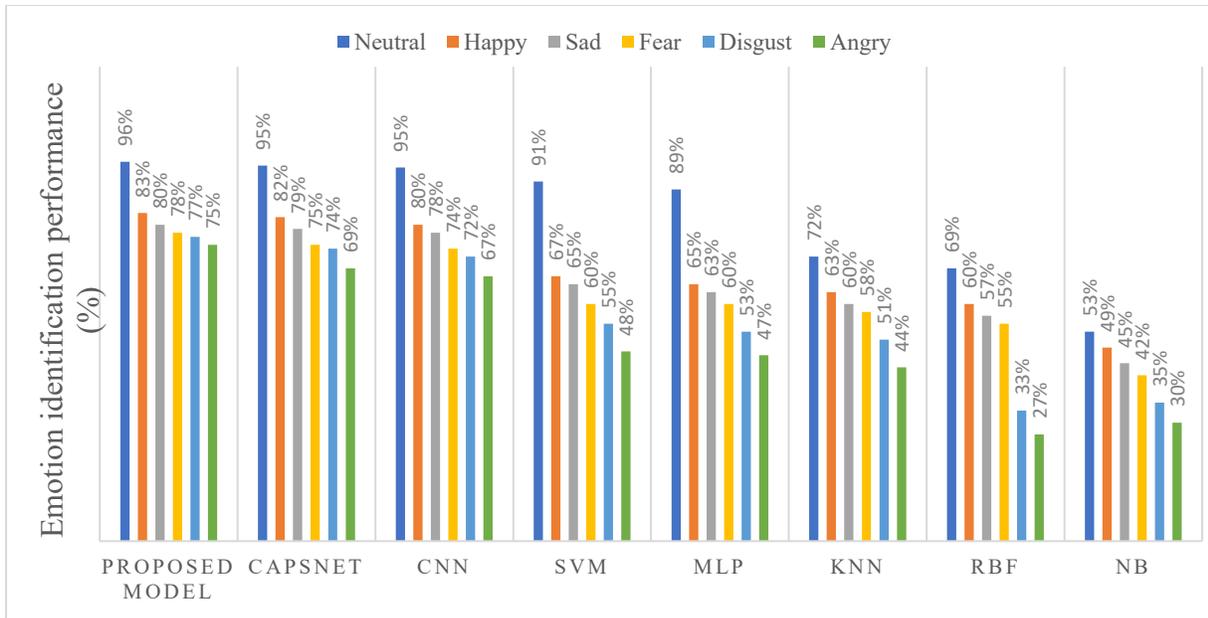

**Fig. 10** Average emotion recognition performance based on the proposed model, CapsNet, CNN, KNN, MLP, RBF, SVM, and NB using CREMA-D corpus

Table 16
Confusion matrix of emotional talking conditions using the CREMA-D database based on the proposed model (%)

| Emotion | Neutral | Happy | Sad | Fear | Disgust | Angry |
|---------|---------|-------|------|------|---------|-------|
| Neutral | **95.6** | 6 | 3 | 4 | 2 | 3.7 |
| Happy | 1.7 | **84.8** | 2 | 2.5 | 1.5 | 4.1 |
| Sad | 1 | 1.5 | **79.8** | 4.3 | 6.1 | 6.5 |
| Fear | 0.5 | 3 | 6 | **78.2** | 7 | 4.2 |
| Disgust | 1 | 2 | 4 | 6 | **78** | 5.9 |
| Angry | 0.2 | 2.7 | 5.2 | 5 | 5.4 | **75.6** |

The four databases that have been utilized to evaluate our proposed model report different emotion recognition accuracies. In SUSAS and Arabic Emirati corpora, the results are homogeneous except for the neutral state, whereas the results of the RAVDESS corpus are homogeneous except for the neutral and angry states. On the other hand, the results of the CREMA-D dataset are heterogeneous. All the datasets prove the superiority of the proposed model, the insufficient results of the NB classifier. Furthermore, the datasets remark the highest accuracies in neutral talking conditions and the lowest accuracies in angry talking conditions.

**Experiment 8: Human listeners**

In order to complete the subjective evaluation of the proposed model utilizing the Emirati-accented database for the assessment of human listeners, the assistance of ten Arabic listeners is required. In this evaluation, a total of 2,160 audio files (10 speakers × 4 sentences × 6 emotional states counting the neutral condition × 9 repetitions/utterance) were utilized. Initially, the ten Arabic listeners were prepared and then were asked to recognize, from the test samples, the right emotion. An illustrative graph on the subjective interpretation of the output measurement by individual listeners is demonstrated in Fig. 11. The graph validates that the performance of human listeners is quite similar to the proposed model performance utilizing the Emirati-accented database.

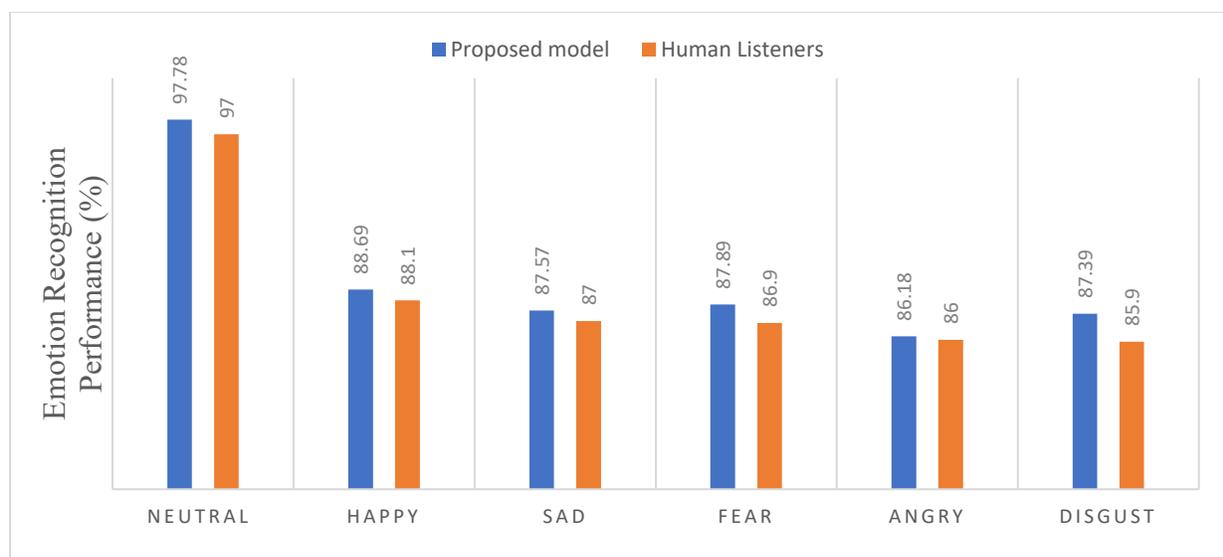

**Fig. 11** Average emotion recognition performance based on human listeners and DC-LSTM COMP-CapsNet using the Arabic Emirati-accented corpus

8) **Experiment 9: Feature extraction techniques**

Five distinct feature extraction techniques are employed in this experiment to assess the superlative technique that is adequate to the proposed model. The five techniques are MFCCs delta-delta, LPC, Hybrid Algorithm DWPD, DCT, and PLDA. The stated results are displayed in Table 17 using the Arabic Emirati-accented database, where MFCC delta-delta is ranked the highest performance amongst the other feature extraction methods. On the other hand, the DCT technique has the least performance. Second optimal techniques are attained by LPC and Hybrid Algorithm DWPD. Pros and cons of each one of the feature extraction techniques are listed in Table 18 (Sunny et al., 2013), (Hibare & Vibhute, 2014; Kaur & Kaur, 2013; Lu & Renals, 2014; Magre et al., 2013; Micallef, 2013; J. Wang et al., 2014).

Table 17

The performance of the proposed work based on five different feature extraction techniques using the Arabic Emirati-accented corpus

| Technique | Average Emotion Recognition Performance |
|---|---|
| MFCCs delta-delta | 89.3% |
| LPC | 80.4% |
| Hybrid Algorithm DWPD | 79.2% |
| PLDA | 75.9% |
| DCT | 66.8% |

Table 18

Advantages and disadvantages of feature extraction techniques

| Method | Pros | Cons |
|---|---|---|
| **MFCCs delta-delta** (Magre et al., 2013) | The human system response can be estimated more tightly than any other system, as the frequency bands are placed logarithmically in MFCC. | In the event of additive noise, MFCC values suffer from low robustness, thus, the values in speech recognition models should be normalized to lessen the impact of noise. |
| **LPC** (Magre et al., 2013) | LPC can decrease the sum of the squared differences between the predicted speech signal and the original speech signal over a finite period. | Since human perceptiveness has fluctuating frequency perceptiveness, the speech recognition system using LPC that approximates constant weighing for the entire spectrum results in hidden outcomes. |
| **Hybrid Algorithm DWPD** (Hibare & Vibhute, 2014)**,** (Sunny et al., 2013) | • Connects the features of both high-frequency components (WPD) and low-frequency components (DWT).<br>• It can not only decrease the high-frequency band to additional segments but also prevent complications in the calculation. | In DWPD, the high frequencies are reduced to get rid of the noise. Yet, from time to time the high-frequency elements may include valuable features of the signal. |
| **PLDA** | It is an adaptable acoustic method that gets use of the variable number of the noncorrelated input frames without any constraints of covariance modeling (Lu & Renals, 2014). | The Gaussian supposition that is on the class conditional distributions, is just a supposition and is not real (J. Wang et al., 2014). |
| **DCT** (Micallef, 2013) | Speed up the model by removing the redundancy from audio data (Kaur & Kaur, 2013). | DCT employs only real-valued, thus, not adaptable. |

## 6. Concluding Remarks

In this work, a novel DC-LSTM COMP-CapsNet has been proposed, applied, and evaluated to enhance "supervised text-independent and speaker-independent emotion recognition performance" using four diverse speech databases. The Dual-Channel LSTM layers are intended to extract sequence-correlated characteristics of Amplitude/Phase signal components and In-phase/Quadrature signal components. Experimental results drawn from this work demonstrate that the proposed model enhances emotion recognition performance. This is due to the ability of our proposed classifier to ensure the compression step and maintain sufficient energy simultaneously along with low computational complexity. Therefore, our proposed model surpasses all other classifiers, which determines that this is the desirable model to be selected as the optimum model, so far, for emotion recognition. The findings of this work illustrate that the proposed work surpasses other conventional classifiers, including MLP, RBF, NB, KNN, SVM, CNN, and the original CapsNet in terms of performance. On the other hand, the running time of training and testing the conventional classifiers is less than that of the proposed architecture. Following the architecture of DC-LSTM COMP-CapsNet, the performance rate is increased from 84.7%, which is the greatest outcome from all of the conventional classifiers (original CapsNet) up to 89.3% based on DC-LSTM COMP-CapsNet.

One of the CapsNet architecture limitations is that it likes to account for everything in the voice being recognized, where the process of CapsNet architecture is done by clutter than that by "orphan" in the dynamic routing process. Therefore, the complexity of the code is huge and may be misleading. The second drawback is that although DC-LSTM COMP-CapsNet is superior to all the conventional classifiers in terms of performance, the complexity of feature extraction and the training running time is more complicated for DC-LSTM COMP-CapsNet architecture. The third limitation is the huge size of the corpus that increases the execution running time complexity. Finally, the proposed model does not give 100% emotion recognition accuracy since datasets contain outliers by nature. This means that the Euclidean distance calculated for two different emotions is the same, which makes it impossible for a classifier to distinguish among such emotions.

Our future work aims to study the novel proposed model for both speaker identification and verification under emotional and stressful talking conditions. Furthermore, we will take advantage of class ties among labels, specifically on multi-label learning issues such as ours. Moreover, we are planning to enhance the performance of the system in an angry state with decreased computational complexity and less running time in the training and testing phases. Ultimately, these outcomes in the capture of emotion-related data can trigger the development of a more robust framework to deal with in-emotion variation simultaneously. Finally, yet importantly, distributed DCT based on MFCC will be utilized and employed rather than only MFCCs delta-delta, resulting in a more robust, faster, and less complex model.

## Acknowledgments
The authors of this work would like to express their gratitude and gratitude to the "University of Sharjah for their assistance through the two competitive research projects entitled Emirati-Accented Speaker and Emotion Recognition Based on Deep Neural Network, No. 19020403139, and Investigation and Analysis of Emirati-Accented Corpus in Neutral and Abnormal Talking Environments for Engineering Applications using Shallow and Deep Classifiers, No. 20020403159."

El Ayadi, M., Kamel, M. S., & Karray, F. (2011). Survey on speech emotion recognition: Features, classification schemes, and databases. *Pattern Recognition*, *44*(3), 572–587. https://doi.org/https://doi.org/10.1016/j.patcog.2010.09.020

Fernández-Diaz, M., & Gallardo-Antolin, A. (2020). An attention Long Short-Term Memory based system for automatic classification of speech intelligibility. *Engineering Applications of Artificial Intelligence*, *96*, 103976.

Goutte, C., & Gaussier, E. (2005). A Probabilistic Interpretation of Precision, Recall and F-Score, with Implication for Evaluation. In D. E. Losada & J. M. Fernández-Luna (Eds.), *Advances in Information Retrieval* (pp. 345–359). Springer Berlin Heidelberg.

Hibare, R., & Vibhute, A. (2014). Feature Extraction Techniques in Speech Processing : A Survey. *International Journal of Computer Applications*, *107*(5). https://doi.org/10.5120/18744-9997

Hogg, R., McKean, J., & Craig, A. (2005). *Introduction to Mathematical Statistics*.

Investopedia. (n.d.). *T-Test Definition*. Retrieved July 5, 2020, from https://www.investopedia.com/terms/t/t-test.asp

Ioffe, S. (2006). Probabilistic Linear Discriminant Analysis. In A. Leonardis, H. Bischof, & A. Pinz (Eds.), *European Conference on Computer Vision* (pp. 531–542). Springer Berlin Heidelberg.

J., L., & I.Tashev. (2015). High-level feature representation using recurrent neural network for speech emotion recognition. *Proc. INTERSPEECH*.

Karandikar, J., McLeay, T., Turner, S., & Schmitz, T. (2015). Tool wear monitoring using naïve Bayes classifiers. *The International Journal of Advanced Manufacturing Technology*, *77*(9), 1613–1626. https://doi.org/10.1007/s00170-014-6560-6

Kaur, S., & Kaur, E. G. (2013). Enhancement of Speech Recognition Algorithm Using DCT and Inverse Wave Transformation. *Journal of Engineering Research and Applications*, *3*(6), 749–754.

Kwabena Patrick, M., Felix Adekoya, A., Abra Mighty, A., & Edward, B. Y. (2019). Capsule Networks – A survey. *Journal of King Saud University - Computer and Information Sciences*. https://doi.org/https://doi.org/10.1016/j.jksuci.2019.09.014

Lin, S.-W., Ying, K.-C., Chen, S.-C., & Lee, Z.-J. (2008). Particle swarm optimization for parameter determination and feature selection of support vector machines. *Expert Systems*
48